\documentclass[%
prb
,twocolumn%
,superscriptaddress
,amsmath,amssymb,aps,nobibnotes,
nofootinbib
,preprintnumbers
]{revtex4-2}
\usepackage[dvipdfmx]{graphicx}
\usepackage{bm}
\usepackage{dcolumn}
\usepackage{color}
\usepackage{tabularx}
\def\urm#1{\scriptstyle{\text{\textrm{\textmd{\textup{#1}}}}}}

\usepackage[normalem]{ulem}  % \sout{old text} for strikeout
\newcolumntype{Y}{>{\centering\arraybackslash}X}
\def\rhom{\overline{\rho}}
\def\rs{r_{\text{s}}}
\def\Exc{E_{\text{xc}}}
\def\exc{\epsilon_{\text{xc}}}

\def\ec{\epsilon_{\text{c}}}
\def\rhou{\rho_{\uparrow}}
\def\rhod{\rho_{\downarrow}}
\begin{document}
\title{Construction of energy density functional for arbitrary spin polarization
  using functional renormalization group}
\author{Takeru Yokota}
\email{takeru.yokota@riken.jp}
\affiliation{Interdisciplinary Theoretical and Mathematical Sciences Program (iTHEMS), RIKEN, Wako, Saitama 351-0198, Japan}
\affiliation{Institute for Solid State Physics, The University of Tokyo, Kashiwa, Chiba 277-8581, Japan}
\author{Tomoya Naito}
\email{tomoya.naito@phys.s.u-tokyo.ac.jp}
\affiliation{Department of Physics, Graduate School of Science, The University of Tokyo, Tokyo 113-0033, Japan}
\affiliation{RIKEN Nishina Center, Wako 351-0198, Japan}
\date{\today}
\preprint{RIKEN-QHP-501}
\preprint{RIKEN-iTHEMS-Report-21}
\begin{abstract}
  We show an application of the functional-renormalization-group aided density functional theory
  to the homogeneous electron gas with arbitrary spin polarization,
  which gives the energy density functional in the
  local spin density approximation.
  The correlation energy per particle
  is calculated at arbitrary Wigner-Seitz radius $ \rs $
  and spin polarization $ \zeta $.
  In the high-density region, our result shows
  good agreement with Monte Carlo (MC) data.
  The agreement with MC data is better in the case of small spin polarization,
  while the discrepancy increases as
  the spin polarization increases.
  The magnetic properties given by our numerical results
  are also discussed.
\end{abstract}
\maketitle
\section{Introduction}
\par
Density functional theory~\cite{
  hoh64,
  koh65,
  koh99}
is a powerful framework for many-body systems employed in various fields,
including condensed matter physics, quantum chemistry, and nuclear physics.
The accuracy of DFT depends on the energy density functional (EDF)~\cite{per01},
which returns the ground-state energy as a functional of the ground-state density.
Although the Hohenberg--Kohn theorem~\cite{hoh64}
guarantees the existence of the \textit{exact} EDF,
it does not provide a microscopic way to derive EDF;
hence its development is a long-standing problem.
\par
A useful framework for the construction of EDF is the effective action formalism \cite{fuk94,fuk95,val97,Furnstahl2020Eur.Phys.J.A56_85}.
In this formalism, Polonyi, Sailer, and Schwenk
\cite{pol02,sch04} put forward an approach
inspired by the functional renormalization group (FRG)~\cite{weg73,wil74,pol84,wet93}.
We refer to this approach as
the functional-renormalization-group aided density functional theory (FRG-DFT).
This approach is based on a functional differential equation, called flow equation,
in a closed form of the effective action,
which corresponds to the free-energy density functional
multiplied by the temperature~\cite{fuk94,fuk95,val97}.
For such a closed form of equation, some systematic approximation methods were proposed~\cite{pol02,kem13,ren15,lia18}.
The FRG-DFT has been numerically applied to
low-dimensional toy models \cite{kem13,ren15,kem17a,lia18,yok18,yok18b,yok21b}
including a mimic of the nuclear systems~\cite{kem17a,yok18,yok18b},
and, recently, applied to electronic systems \cite{yok19,yok21},
where we achieved the first application to the two and three dimensional cases in the study of the homogeneous electron gas (HEG)
and EDFs for electrons are constructed in the local density approximation.
\par
In the previous works of the FRG-DFT~\cite{pol02,sch04,kem13,ren15,kem17a,lia18,yok18,yok18b,yok19,yok21,yok21b},
the EDF of total particle-number density has been studied.
The Hohenberg--Kohn theorem guarantees that the ground-state properties are determined by the total particle-number density.
However, the Kohn--Sham scheme~\cite{koh65}, which is a commonly used method in DFT,
has limited power when formulated based on the EDF of total particle number:
Although it is suitable for accurate analysis of the ground-state energy and density,
the calculations of other quantities are not always straightforward~\cite{Martin2004_CambridgeUniversityPress}.
A way to extend Kohn--Sham scheme is introducing
EDF depending on additional densities~\cite{
  Jansen1991Phys.Rev.B43_12025};
one of such generalizations is taking into account
the particle-number densities of each spin component
to describe the magnetic properties~\cite{Barth1972J.Phys.C5_1629}.
Such a formulation of EDF for multi-component systems
can also be directly applied to analyses of nuclear matter,
which is infinite system of nucleons composed of various components
such as protons, neutrons, and
hyperons~\cite{
  Akmal1998Phys.Rev.C58_1804,
  Dickhoff2004Prog.Part.Nucl.Phys.52_377,
  Stone2007Prog.Part.Nucl.Phys.58_587,
  Gandolfi2010Mon.Not.R.Astron.Soc.404_L35,
  lat12,
  Togashi2013Nucl.Phys.A902_53,
  Togashi2016Phys.Rev.C93_035808,
  RevModPhys.89.015007,
  Tong2018Phys.Rev.C98_054302,
  Myo2019Phys.Rev.C99_024312,
  Wang2021Phys.Rev.C103_054319}.
In the context of the FRG-DFT, the EDF depending
on the pairing density in addition to the particle-number density
to describe superfluid systems
was discussed in Ref.~\cite{yok20}.
\par
As for electron systems,
a simple EDF of the particle-number densities of each spin component
is that in the local spin density approximation (LSDA),
in which the exchange--correlation part of the EDF
$ \Exc \left[ \rhou, \rhod \right] $
is approximated as
\begin{equation}
  \label{eq:lsdaedf}
  \Exc \left[ \rhou, \rhod \right]
  \approx
  \int
  d \bm{x}
  \,
  \left(
    \rhou \left( \bm{x} \right)
    +
    \rhod \left( \bm{x} \right)
  \right)
  \exc
  \left( \rhou \left( \bm{x} \right), \rhod \left( \bm{x} \right) \right),
\end{equation}
where $ \exc \left( \rhou, \rhod \right) $
is the exchange--correlation energy per particle, i.e., energy density, of the HEG of densities of particles 
with up spin $ \rhou $ and down spin $ \rhod $.
At some values of $ \rhou $ and $ \rhod $,
$ \exc \left( \rhou, \rhod \right) $
has been obtained by the quantum Monte Carlo (QMC) calculations~\cite{loo16}.
The QMC results are of significance
not only for the construction of EDF~\cite{cep80, vos80, per81}
but also for the discussion of the properties of the HEG itself:
In three dimensions,
the existence of the phase transitions
from paramagnetic phase to partially polarized or ferromagnetic phase and
from ferromagnetic phase to Wigner crystal is predicted~\cite{cep78,cep80,ort94,ort97,kwo98,ort99,zon02,dru04,spi13},
although the predicted values of the transition densities are not always similar among these works.
In two dimensions,
the existence of the ferromagnetic phase is indicated by Refs.~\cite{tan89,kwo93,rap96,att02,att03,gor03},
while it is suggested in Ref.~\cite{dru09}
that the ferromagnetic phase is never stable compared to the paramagnetic one.
\par
Since the QMC results are available only at few values of the densities due to the numerical cost,
an empirical fitting function for the data is needed to obtain the value of EDF at arbitrary densities.
In our previous work \cite{yok21},
we applied the FRG-DFT for the calculation of the correlation energy per particle
in the spin-unpolarized (paramagnetic) case.
We obtained the values at many values of density, namely
65536 points in the Wigner-Seitz radii
$ \rs \in \left[ 10^{-6} \, \mathrm{a.u.}, 100 \, \mathrm{a.u.} \right) $,
which enables us to determine EDF almost without the dependence on the fitting function.
\par                                       
In this paper, we extend our works~\cite{yok19,yok21} to the case of arbitrary spin polarization
$ \zeta = \left( \rhou - \rhod \right) / \left( \rhou + \rhod \right) $.
We give a formulation of the FRG-DFT with arbitrary spin polarization and dimensions.
In the vertex expansion, which is a functional Taylor expansion around densities of interest,
we give the flow equations for the density correlation functions at arbitrary order.
The correlation energy per particle $ \ec \left( \rs, \zeta \right) $
is calculated at various $ \rs $ and $ \zeta $
in the three- and two-dimensional cases.
In high-density region, our result shows good agreement with QMC data
as a result of the fact that our correlation energy reproduces the exact behavior at high density given by the Gell-Mann--Brueckner (GB) resummation~\cite{gel57,raj77}.
The agreement with QMC data is better in the case of small spin polarization,
while the discrepancy increases as the spin polarization increases,
which causes the absence of the magnetic phase transition predicted by the QMC calculations.
\par
We also discuss the interpolation function
$ f_{\text{c}} \left( \rs, \zeta \right) $
defined by
\begin{equation}
  \label{eq:fc}
  f_{\text{c}} \left(\rs, \zeta \right)
  =
  \frac{
    \ec \left( \rs, \zeta \right)
    -
    \ec \left( \rs, 0 \right)}
  {\ec \left( \rs, 1 \right) - \ec \left( \rs, 0 \right)},
\end{equation}
which characterizes the $ \zeta $ dependence of EDF.
Since $ \ec $ of only a few points $ \zeta $ and $ \rs $ are available in QMC calculation,
interpolation with respect to $ \zeta $ is demanded to obtain $ \ec $ for arbitrary $ \zeta $
to perform LSDA calculations,
as similar to the interpolation with respect to $ \rs $.
A popular approach to determine $ \ec \left( \rs, \zeta \right) $
\cite{
  Barth1972J.Phys.C5_1629,
  Martin2004_CambridgeUniversityPress}
is the approximation
$ f_{\text{c}} \left(\rs, \zeta \right) \approx f_{\text{x}} \left( \zeta \right) $
with the interpolation function for the exchange part
$ f_{\text{x}} \left( \zeta \right) $
given analytically.
We find the deviation of $ f_{\text{c}} \left( \rs, \zeta \right) $
from $ f_{\text{x}} \left( \zeta \right) $ at small $ \rs $,
where the FRG-DFT gives accurate results.
\par
This paper is organized as follows:
In Sec.~\ref{sec:form}, we shall show an FRG-DFT formalism with arbitrary spin polarization and dimensions.
The flow equations for the density correlation functions at arbitrary order are given.
The expression for the correlation energy density is given in the case of the second-order truncation.
It shall be analytically shown that the expression reproduces the GB resummation at high density.
Our numerical results are presented in Sec.~\ref{sec:result}.
We show the result of the $ \rs $ and $ \zeta $ dependences of the correlation energy
and discuss the magnetic properties given by our calculation.
Section \ref{sec:conc} is devoted to the conclusion.
In Appendix~\ref{sec:deriv}, we give the derivation of Eqs.~\eqref{eq:flow0} and \eqref{eq:flown}.
\section{Formalism}
\label{sec:form}
\par
We consider the HEG with the homogeneous density $ \rhom $ and the inverse temperature $ \beta $
neutralized by the background ions with the same density.
In order to apply our formalism to the three- and two-dimensional cases,
we present the formulation in general spatial dimensions $ D $.
The action in the imaginary time formalism is given by
\begin{subequations}
  \begin{align}
    S \left[ \psi, \psi^* \right]
    & =
      S_{\text{el}}  \left[ \psi, \psi^* \right]
      +
      S_{\text{int}} \left[ \psi, \psi^* \right],
      \label{eq:acti} \\
    S_{\text{el}}  \left[ \psi, \psi^* \right]
    & =
      \sum_s
      \int_X
      \psi_s^* \left( X_{\epsilon} \right)
      \left(
      \partial_{\tau}
      -
      \frac{1}{2}
%      \sout{\Delta}
      \nabla^2
      \right)
      \psi_s \left( X \right) ,
      \notag \\
    S_{\text{int}} \left[ \psi, \psi^* \right]
    & =
      \frac{1}{2}
      \iint_{X, \, X'}
      U_{\urm{2b}} \left( X, X' \right)
      \hat{\rho}_{\Delta} \left( X \right)
      \hat{\rho}_{\Delta} \left( X' \right).
  \end{align}
\end{subequations}
Here, we have introduced the following notations:
$ X = \left( \tau, \bm{x} \right) $
with the imaginary time $ \tau $ and spatial coordinate $ \bm{x} $,
the coordinate integral $ \int_X = \int_0^{\beta} d \tau \int d \bm{x} $,
$ \hat{\rho}_{\Delta} \left( X \right) = \sum_s \hat{\rho}_s \left( X \right) - \rhom $
with the electron field $ \psi_s \left( X \right) $ having spin $ s $ and
the electron density operator
$ \hat{\rho}_s \left( X \right) = \psi_s^* \left( X_{\epsilon} \right) \psi_s \left( X \right) $,
and the simultaneous Coulomb interaction
\begin{equation}
  \label{eq:Coul}
  U_{\text{2b}} \left( X, X' \right)
  =
  \frac{1}{\left| \bm{x} - \bm{x}' \right|}
  \delta \left( \tau - \tau' \right).
\end{equation}
The interaction term $ S_{\text{int}} $
contains not only the electron-electron interaction term
but also the electron-ion and ion-ion ones,
which cancel each other, and consequently avoid the divergence from Hartree term
in the infinite system~\cite{yok19}.
Additionally, $ X_{\epsilon} = \left( \tau + \epsilon, \bm{x} \right) $
with a positive infinitesimal $ \epsilon $
has been introduced so that the corresponding Hamiltonian is normal ordered.
\subsection{FRG-DFT flow equation}
\par
Following Refs.~\cite{pol02,sch04},
we analyze the evolution of the system
with respect to the gradual change of the strength of the two-body Coulomb interaction [Eq.~\eqref{eq:Coul}].
For this purpose, the evolution parameter $ \lambda \in \left[ 0, 1 \right] $
is attached
to the interaction term $ S_{\text{int}} \left[ \psi, \psi^* \right] $ as follows:
\begin{align}
  S_{\lambda} \left[ \psi, \psi^* \right]
  =
  S_{\text{el}} \left[ \psi, \psi^* \right]
  +
  \lambda
  S_{\text{int}} \left[ \psi, \psi^* \right] .
\end{align}
This action becomes that for a non-interacting system at $ \lambda = 0 $
and Eq.~\eqref{eq:acti} at $ \lambda = 1 $.
\par
Our previous works~\cite{yok19,yok21}
were focused on EDFs of the total density.
In the present study, we extend our analysis
to the case of EDFs for spin polarized systems.
For this purpose,
%\sout{For the purpose of the derivation of EDF of densities of each spin component,}
we introduce the generating functional for the correlation functions for
$ \hat{\rho}_{{\uparrow}, \, {\downarrow}} \left( X \right) $:
\begin{equation}
  Z_{\lambda} \left[ J_{\uparrow}, J_{\downarrow} \right]
  =
  \iint
  \mathcal{D} \psi \,
  \mathcal{D} \psi^* \,
  e^{- S_{\lambda} \left[ \psi, \psi^* \right]
    +
    \sum_{s}
    \int_X
    \hat{\rho}_s \left( X \right)
    J_s \left( X \right)}.
\end{equation}
The Legendre transformation of the generating functional for connected correlation functions
$ W_{\lambda} \left[ J_{\uparrow}, J_{\downarrow} \right]
= \ln Z_{\lambda} \left[ J_{\uparrow}, J_{\downarrow} \right] $
gives the effective action:
\begin{align}
  \label{eq:gamma}
  \Gamma_{\lambda} \left[ \rhou, \rhod \right]
  = & \,
      \sup_{J_{\uparrow}, \, J_{\downarrow}}
      \left(
      \sum_s
      \int_X
      \rho_s \left( X \right)
      J_s \left( X \right)
      -
      W_{\lambda} \left[ J_{\uparrow}, J_{\downarrow} \right]
      \right)
      \notag \\
  = & \,
      \sum_s
      \int_X
      \rho_s \left( X \right)
      J_{\text{sup}, \, \lambda, \, s}
      \left[ \rhou, \rhod \right] \left( X \right)
      \notag \\
    & \,
      -
      W_{\lambda}
      \left[
      J_{\text{sup}, \, \lambda, \, {\uparrow}} \left[ \rhou, \rhod \right],
      J_{\text{sup}, \, \lambda, \, {\downarrow}} \left[ \rhou, \rhod \right]
      \right].
\end{align}
Here,
the external field
% $ J_{\text{sup}, \, \lambda, \, {\uparrow}} \left[ \rhou, \rhod \right] $,
$ J_{\text{sup}, \, \lambda, \, s} \left[ \rhou, \rhod \right] $,
which gives the supremum of the first line of Eq.~\eqref{eq:gamma},
satisfies
\begin{equation}
  \label{eq:jsup}
  \frac{\delta W_{\lambda}
    \left[
      J_{\text{sup}, \, \lambda, \, {\uparrow}} \left[ \rhou, \rhod \right],
      J_{\text{sup}, \, \lambda, \, {\downarrow}} \left[ \rhou, \rhod \right]
    \right]}
  {\delta J_s \left( X \right)}
  =
  \rho_s \left( X \right).
\end{equation}
The equilibrium densities $ \rho_{\text{eq}, \, s} \left( X \right) $
are determined through
\begin{equation}
  \frac{\delta \Gamma_{\lambda} \left[ \rho_{\text{eq}, \, {\uparrow}}, \rho_{\text{eq}, \, {\downarrow}} \right]}
  {\delta \rho_s \left( X \right)}
  =
  \mu_s,
\end{equation}
where $ \mu_s $ is
the chemical potential of the electrons with the spin component $ s $.
In this case, Eq.~\eqref{eq:gamma} becomes
\begin{equation}
  \label{eq:gamma_eq}
  \Gamma_{\lambda}
  \left[ \rho_{\text{eq}, \, {\uparrow}}, \rho_{\text{eq}, \, {\downarrow}} \right]
  =
  \sum_s
  \mu_s
  \int_X
  \rho_{\text{eq}, \, s} \left( X \right)
  -
  W_{\lambda} \left[ \mu_{\uparrow}, \mu_{\downarrow} \right],
\end{equation}
because of
$ J_{\text{sup}, \, \lambda, \, s} \left[ \rho_{\text{eq}, \, {\uparrow}}, \rho_{\text{eq}, \, {\downarrow}} \right] \left( X \right) = \mu_s $,
which is obtained by the relation
\begin{equation}
  \label{eq:gamma1}
  \frac{\delta \Gamma_{\lambda} \left[ \rhou, \rhod \right]}{\delta \rho_s \left( X \right)}
  =
  J_{\text{sup}, \, \lambda, \, s} \left[ \rhou, \rhod \right] \left( X \right).
\end{equation}
Since $ W_{\lambda} \left[ \mu_{\uparrow}, \mu_{\downarrow} \right] / \beta $
is the grand potential,
$ \Gamma_{\lambda} \left[ \rho_{\text{eq}, \, {\uparrow}}, \rho_{\text{eq}, \, {\downarrow}} \right] $
gives the free energy multiplied by $ \beta $. % from Eq.~\eqref{eq:gamma_eq}.
At zero temperature,
$ \rho_{\text{eq}, \, s} \left( X \right) $ and
$ \beta^{-1} \Gamma_{\lambda} \left[ \rho_{\text{eq}, \, {\uparrow}}, \rho_{\text{eq}, \, {\downarrow}} \right] $
are reduced to the ground-state density and energy, respectively.
Therefore, the EDF of densities with each spin component is given 
%\sout{by} 
as follows \cite{kem13}:
\begin{equation}
  \label{eq:edf}
  E_{\lambda} \left[ \rhou, \rhod \right]
  =
  \lim_{\beta \to \infty}
  \beta^{-1} \Gamma_{\lambda} \left[ \rhou, \rhod \right],
\end{equation}
which is a natural extension of the relation
between EDF of total density and the effective action~\cite{fuk94,val97}.
\par
The evolution of $ \Gamma_{\lambda} \left[ \rhou, \rhod \right] $
can be described by a functional differential equation.
This equation is derived in the same manner as EDF
for total density~\cite{sch04,kem13,kem17a,yok18,yok19} and reads
\begin{widetext}
  \begin{equation}
    \label{eq:flow_eq}
    \partial_{\lambda}
    \Gamma_{\lambda} \left[ \rhou, \rhod \right]
    =
    \frac{1}{2}
    \iint_{X, \, X'}
    U_{\text{2b}} \left( X - X' \right)
    \left[
      \rho_{\Delta} \left( X \right)
      \rho_{\Delta} \left( X' \right)
      +
      \sum_{s, \, s'}
      \Gamma^{\text{($ 2 $)} {-1}}_{\lambda, \, s s'} \left[ \rhou, \rhod \right]
      \left( X_{\epsilon'}, X' \right)
      -
      \sum_s
      \rho_s \left( X \right)
      \delta \left( \bm{x} - \bm{x}' \right)
    \right].
  \end{equation}
\end{widetext}
Here, we have introduced
$ \rho_{\Delta} \left( X \right) = \sum_s \rho_s \left( X \right) - \rhom $
and the inverse of the second derivative of the effective action
$ \Gamma^{\text{($ 2 $)} {-1}}_{\lambda, \, s s'} \left[ \rhou, \rhod \right] \left( X, X' \right) $
defined through the following relation:
\begin{align}
  & \sum_{s''}
    \int_{X''}
    \Gamma^{\text{($ 2 $)} {-1}}_{\lambda, \, s s''} \left[ \rhou, \rhod \right] \left( X, X'' \right)
    \frac{\delta^2 \Gamma_{\lambda} \left[ \rhou, \rhod \right]}
    {\delta \rho_{s''} \left( X'' \right) \, \delta \rho_{s'} \left( X' \right)}
    \notag \\
  & =
    \delta_{s s'} \delta \left( X - X' \right),
\end{align}
where
$ \delta \left( X - X' \right) = \delta \left( \tau - \tau' \right) \delta \left( \bm{x} - \bm{x}' \right) $
with $ X = \left( \tau, \bm{x} \right) $ and $ X' = \left( \tau', \bm{x}' \right) $.
\subsection{Vertex expansion}
\par
%\begin{figure*}[tb]
%  \centering
%	\begin{alignat*}{3}
%	\mathrm{(a)}
%	\qquad
%	&
%	\partial_\lambda\Gamma_\lambda
%	=
%	\frac{1}{2}
%	\parbox[c]{8em}{\includegraphics[width=8em]{fig_fdiagram1.pdf}}
%	+
%	\frac{1}{2}
%	\parbox[c]{5em}{\includegraphics[width=5em]{fig_fdiagram2.pdf}}
%	-
%	\frac{1}{2}
%	\parbox[c]{5em}{\includegraphics[width=5em]{fig_fdiagram3.pdf}}
%	\\
%	\mathrm{(b)}
%	\qquad
%	&
%	\partial_\lambda G^{(m)}_{\lambda,s_1\cdots s_m}
%	(X_1,\ldots,X_m)
%	=
%	\parbox[c]{9em}{\includegraphics[width=9em]{fig_fdiagram4.pdf}}
%	-
%	\parbox[c]{10em}{\includegraphics[width=10em]{fig_fdiagram5.pdf}}
%	-
%	\frac{1}{2}
%	\parbox[c]{7em}{\includegraphics[width=7em]{fig_fdiagram6.pdf}}
%	+
%	\frac{1}{2}
%	\parbox[c]{8em}{\includegraphics[width=8em]{fig_fdiagram7.pdf}}
%	\\
%	&-\frac{1}{2}
%	\end{alignat*}
%  \caption{\cor{Diagrammatic representation of Eqs.~\eqref{eq:flow0} and \eqref{eq:flown}. The wavy lines represent the two-body interaction $U_{\rm 2b}(X-X')$.
%  The numbers $1,\, 2,\ldots,\, m$ attached to the diagrams of the correlation functions 
%  $G^{(m+2)}_\lambda,\,G^{(m+1)}_\lambda,\ldots$ stand for the
%  sets of the coordinate and spin $(X_1,s_1),\, (X_2,s_2),\ldots, (X_m, s_m)$.}}
%  \label{fig:fdiagram}
%\end{figure*}
\begin{figure*}[tb]
  \centering
  \includegraphics[width=2\columnwidth]{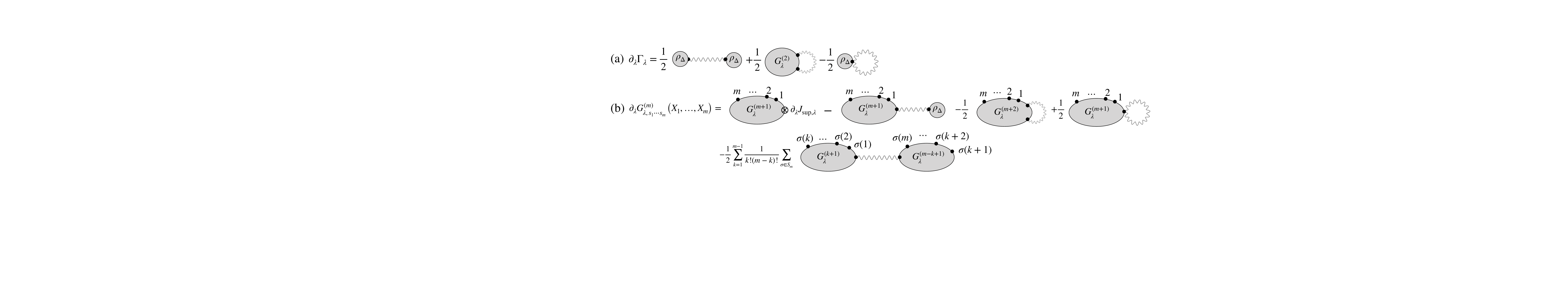}
  \caption{
  Diagrammatic representation of Eqs.~\eqref{eq:flow0} and \eqref{eq:flown}. The wavy lines represent the two-body interaction $ U_{\text{2b}} \left( X - X' \right) $.
  The numbers $ 1 $, $ 2 $, \ldots, $ m $ attached to the diagrams of the correlation functions
  $ G^{\text{($ m+2 $)}}_{\lambda} $, $ G^{\text{($ m+1 $)}}_{\lambda} $, \ldots 
  stand for the
  sets of the coordinate and spin
  $ \left( X_1, s_1 \right) $, $ \left( X_2, s_2 \right) $, \ldots, $ \left( X_m, s_m \right) $.
  }
  \label{fig:fdiagram}
\end{figure*}
In principle, $ \Gamma_{\lambda} \left[ \rhou, \rhod \right] $
is obtained by solving Eq.~\eqref{eq:flow_eq}
with using the non-interacting system as the initial condition.
In practice, however, some approximation for
$ \Gamma_{\lambda} \left[ \rhou, \rhod \right] $ is needed to solve Eq.~\eqref{eq:flow_eq}
due to difficulty of direct numerical treatment of functional differential equations.
Following Refs.~\cite{pol02,sch04}, we introduce the vertex expansion scheme,
where a functional Taylor expansion at some densities of interest
is applied and the differential equations for the Taylor coefficients of
$ \Gamma_{\lambda} \left[ \rhou, \rhod \right] $ are truncated at some order.
As derived in Appendix~\ref{sec:deriv},
these differential equations up to an arbitrary order
read in terms of the density correlation functions as follows:
\begin{widetext}
  \begin{align}
    & \partial_{\lambda}
      \Gamma_{\lambda} \left[ \rhou, \rhod \right]
      =
      \frac{1}{2}
      \iint_{X, \, X'}
      U_{\text{2b}} \left( X - X' \right)
      \left[
      \rho_{\Delta} \left( X \right)
      \rho_{\Delta} \left( X' \right)
      +
      \sum_{s, \, s'}
      G^{\text{($ 2 $)}}_{\lambda, \, s s'} \left( X_{\epsilon'}, X' \right)
      -
      \sum_s
      \rho_s \left( X \right)
      \delta \left( \bm{x} - \bm{x}' \right)
      \right],
      \label{eq:flow0} \\
    & \partial_{\lambda}
      G^{\text{($ m $)}}_{\lambda, \, s_1 \cdots s_m} \left( X_1, \ldots , X_m \right)
      \notag \\
    & =
      \sum_s
      \int_X
      G^{\text{($ m + 1 $)}}_{\lambda, \, s, s_1 \cdots s_m} \left( X, X_1, \ldots , X_m \right)
      \partial_{\lambda}
      J_{\text{sup}, \, \lambda, \, s} \left[ \rhou, \rhod \right] \left( X \right)
      \notag \\
    & \quad
      -
      \frac{1}{2}
      \sum_s
      \iint_{X, \, X'}
      U_{\text{2b}} \left( X, X' \right)
      \left(
      \rho_{\Delta} \left( X \right)
      G^{\text{($ m + 1 $)}}_{\lambda, \, s s_1 \cdots s_m} \left( X', X_1, \ldots , X_m \right)
      +
      \rho_{\Delta} \left( X' \right)
      G^{\text{($ m + 1 $)}}_{\lambda, \, s s_1 \cdots s_m} \left( X, X_1, \ldots , X_m \right)
      \right)
      \notag \\	
    & \quad
      -
      \frac{1}{2}
      \sum_{s, \, s'}
      \iint_{X, \, X'}
      U_{\text{2b}} \left( X, X' \right)
      \left(
      G^{\text{($ m + 2 $)}}_{\lambda, \, s s' s_1 \cdots s_m} \left( X_{\epsilon'}, X', X_1, \ldots , X_m \right)
      -
      G^{\text{($ m + 1 $)}}_{\lambda, \, s s_1 \cdots s_m} \left( X, X_1, \ldots , X_m \right)
      \delta_{s s'}
      \delta \left( \bm{x} - \bm{x}' \right)
      \vphantom{\sum_{k = 1}^{m - 1}}
      \right.
      \notag \\
    & \qquad
      \left.
      +
      \sum_{k = 1}^{m - 1}
      \frac{1}{k! \, \left( m - k \right)!}
      \sum_{\sigma \in S_m}
      G^{\text{($ k + 1 $)}}_{\lambda, \, s s_{\sigma \left( 1 \right)} \cdots s_{\sigma \left( k \right)}}
      \left( X, X_{\sigma \left( 1 \right)}, \ldots , X_{\sigma \left( k \right)} \right)
      G^{\text{($ m - k + 1 $)}}_{\lambda, \, s' s_{\sigma \left( k + 1 \right)} \cdots s_{\sigma \left( m \right)}}
      \left( X', X_{\sigma \left( k + 1 \right)}, \ldots , X_{\sigma \left( m \right)} \right)
      \right),
      \label{eq:flown}
  \end{align}
\end{widetext}
where $ m \geq 1 $ is an integer, $ S_m $ stands for the symmetry group of order $ m $,
and the $ m $-point density correlation function is defined by
\begin{align}
  \label{eq:gdef}
  & G^{\text{($ m $)}}_{\lambda, \, s_1 \cdots s_m} \left( X_1, \ldots, X_m \right)
    \notag \\
  & =
    \frac{\delta^m W_{\lambda} \left[
    J_{\text{sup}, \, \lambda, \, {\uparrow}} \left[ \rhou, \rhod \right],
    J_{\text{sup}, \, \lambda, \, {\downarrow}} \left[ \rhou, \rhod \right] \right]}
    {\delta J_{s_1} \left( X_1 \right) \ldots \delta J_{s_m} \left( X_m \right)}.
\end{align}
%\cor{
The diagrammatic representation of Eqs.~\eqref{eq:flow0}
and \eqref{eq:flown} is given in Fig.~\ref{fig:fdiagram}.
%}
Note that Eq.~\eqref{eq:flown} at $ m = 1 $ should be regarded as
an equation describing the evolution of
$ J_{\text{sup}, \, \lambda, \, s} \left[ \rhou, \rhod \right] \left( X \right) $
since the left-hand side is already determined by
$ \partial_{\lambda} G^{\text{($ 1 $)}}_{\lambda, \, s_1} \left( X_1 \right) 
= \partial_{\lambda} \rho_{s_1} \left( X_1 \right) = 0 $ as
obtained from Eq.~\eqref{eq:jsup}.
\subsection{Application to LSDA}
\par
Our purpose is the application to the LSDA EDF,
which is described in terms of the effective action by
\begin{align}
  & \Gamma_{\text{xc}, \, \lambda} \left[ \rhou, \rhod \right]
    \notag \\
  & \approx
    \int_X
    \left( \rhou \left( X \right) + \rhod \left( X \right) \right)
    \epsilon_{\text{xc}, \, \lambda}
    \left( \rhou \left( X \right), \rhod \left( X \right) \right).
    \label{eq:lsda}
\end{align}
Here,
$ \Gamma_{\text{xc}, \, \lambda} \left[ \rhou, \rhod \right] $
is given by the exchange--correlation part of the effective action
\begin{align*}
  \Gamma_{\text{xc}, \, \lambda} \left[ \rhou, \rhod \right]
  = & \,
      \Gamma_{\lambda} \left[ \rhou, \rhod \right]
      -
      \Gamma_{\lambda = 0} \left[ \rhou, \rhod \right]
      \notag \\
    & \,
      -
      \frac{1}{2}
      \iint_{X, \, X'}
      \lambda
      U_{\text{2b}} \left( X - X' \right)
      \rho_{\Delta} \left( X \right)
      \rho_{\Delta} \left( X' \right)
\end{align*}
with the second and third terms of the right-hand side
corresponding to the kinetic and Hartree terms, respectively, and
$ \epsilon_{\text{xc}, \, \lambda} \left( \rhou, \rhod \right) $
is the exchange--correlation energy per particle obtained in the homogeneous case with
densities $ \rhou $ and $ \rhod $.
Actually, Eq.~\eqref{eq:lsda} is reduced to the conventional definition of LSDA for EDF
Eq.~\eqref{eq:lsdaedf} as obtained from Eq.~\eqref{eq:edf}
with the density independent of the imaginary time
$ \rho_s \left( X \right) = \rho_s \left( \bm{x} \right) $.
By putting Eq.~\eqref{eq:lsda} into Eq.~\eqref{eq:flow0},
we have an equation to determine
$ \epsilon_{\text{xc}, \, \lambda} \left( \rhou, \rhod \right) $:
\begin{align}
  & \int_X
    \left( \rhou \left( X \right) + \rhod \left( X \right) \right)
    \partial_{\lambda}
    \epsilon_{\text{xc}, \, \lambda} \left( \rhou \left( X \right), \rhod \left( X \right) \right)
    \notag \\
  & \approx
    \frac{1}{2}
    \iint_{X, \, X'}
    U_{\text{2b}} \left( X - X' \right)
    \notag \\
  & \quad
    \times
    \left[
    \sum_{s, \, s'}
    G^{\text{($ 2 $)}}_{\lambda,s s'} \left( X_{\epsilon'}, X' \right)
    -
    \sum_s
    \rho_s \left( X \right)
    \delta \left( \bm{x} - \bm{x}' \right)
    \right] .
    \label{eq:xcinhom}
\end{align}
Particularly, Eq.~\eqref{eq:xcinhom} becomes exact at the homogeneous limit~\cite{lew19}.
Given the homogeneous densities $ \rhom_s $
satisfying $ \rhom_{\uparrow} + \rhom_{\downarrow} = \rhom $,
Eq.~\eqref{eq:xcinhom} is reduced to
\begin{align}
  & \partial_{\lambda}
    \epsilon_{\text{xc}, \, \lambda}
    \left( \rhom_{\uparrow}, \rhom_{\downarrow} \right)
    \notag \\
  & =
    \frac{1}{2 \rhom}
    \int_X
    U_{\text{2b}} \left( X \right)
    \left[
    \sum_{s, \, s'}
    G^{\text{($ 2 $)}}_{\lambda, \, s s'} \left( X_{\epsilon'}, 0 \right)
    -
    \rhom
    \delta \left( \bm{x} \right)
    \right].
    \label{eq:exc}
\end{align}
Here, we have used
$ G^{\text{($ 2 $)}}_{\lambda, \, s s'} \left( X_{\epsilon'}, X' \right)
=
G^{\text{($ 2 $)}}_{\lambda, \, s s'} \left( X_{\epsilon'} - X', 0 \right)$,
which follows from the translational symmetry.
Equation~\eqref{eq:exc} is further reduced to the equation for the correlation part
$ \epsilon_{\text{c}, \, \lambda} \left( \rhom_{\uparrow}, \rhom_{\downarrow} \right)
=
\epsilon_{\text{xc}, \, \lambda} \left( \rhom_{\uparrow}, \rhom_{\downarrow} \right)
-
\epsilon_{\text{x}, \, \lambda} \left( \rhom_{\uparrow}, \rhom_{\downarrow} \right) $:
\begin{align}
  & \partial_{\lambda}
    \epsilon_{\text{c}, \, \lambda}
    \left( \rhom_{\uparrow}, \rhom_{\downarrow} \right)
    \notag \\
  & =
    \frac{1}{2 \rhom}
    \sum_{s, \, s'}
    \int_X
    U_{\text{2b}} \left( X \right)
    \left[
    G^{\text{($ 2 $)}}_{\lambda, \, s s'} \left( X_{\epsilon'}, 0 \right)
    -
    G^{\text{($ 2 $)}}_{\lambda=0, s s'} \left( X_{\epsilon'}, 0 \right)
    \right],
    \label{eq:ec}
\end{align}
which is obtained by use of the expression for the exchange part
$ \epsilon_{\text{x}, \, \lambda} \left( \rhom_{\uparrow}, \rhom_{\downarrow} \right) $:
\begin{align}
  & \epsilon_{\text{x}, \, \lambda} \left( \rhom_{\uparrow}, \rhom_{\downarrow} \right)
    \notag \\
  & =
    \frac{\lambda}{2 \rhom}
    \int_X
    U_{\text{2b}} \left( X \right)
    \left[
    \sum_{s, \, s'}
    G^{\text{($ 2 $)}}_{\lambda = 0, \, s s'} \left( X_{\epsilon'}, 0 \right)
    -
    \rhom \delta \left( \bm{x} \right)
    \right].
\end{align}
The analytic result of
$ \epsilon_{\text{x}, \, \lambda} \left( \rhom_{\uparrow}, \rhom_{\downarrow} \right) $
is known as follows~\cite{dir30,fri97}:
\begin{equation}
  \epsilon_{\text{x}, \, \lambda} \left( \rhom_{\uparrow}, \rhom_{\downarrow} \right)
  =
  -
  \frac{\lambda a_D}{\rs}
  \left[
    1 + \left( 2^{1/D} - 1 \right)
    f_{\text{x}} \left( \zeta \right)
  \right].
  \label{eq:ex}
\end{equation}
Here, $ \zeta = \left( \rhom_{\uparrow} - \rhom_{\downarrow} \right) / \rhom $
is the spin polarization and $ \rs $ is the Wigner--Seitz radius
given by $ \rs = \left( V_D \rhom \right)^{- 1 / D} $
with $ V_{D = 2} = \pi $ and $ V_{D = 3} = 4 \pi /3 $ being the volume of a $ D $-dimensional unit sphere.
The interpolation function $ f_{\text{x}} \left( \zeta \right) $ is defined by
\begin{equation}
  f_{\text{x}} \left( \zeta \right)
  =
  \frac{
    \left( 1 + \zeta \right)^{\left( D + 1 \right) / D}
    +
    \left( 1 - \zeta \right)^{\left( D + 1 \right) / D}-2}
  {2^{\left( D + 1 \right) / D} - 2}.
  \label{eq:fx}
\end{equation}
The coefficient $ a_D $ is given by
$ a_2 = 4 \sqrt{2} / \left( 3 \pi \right) $
and
$ a_3 = 3 \left[ 3 / \left(16\pi\right) \right]^{2/3} $.
The momentum representation is a convenient choice 
for the homogeneous case.
Then, Eqs.~\eqref{eq:flown} and \eqref{eq:ec} are written as follows:
\begin{widetext}
  \begin{align}
    & \partial_{\lambda}
      \epsilon_{\text{c}, \, \lambda} \left( \rhom_{\uparrow}, \rhom_{\downarrow} \right)
      =
      \frac{1}{2 \rhom}
      \sum_{s, \, s'}
      \int_P
      \tilde{U} \left( \bm{p} \right)
      e^{i P^0 \epsilon'}
      \left[
      \tilde{G}^{\text{($ 2 $)}}_{\lambda, \, s s'} \left( P \right)
      -
      \tilde{G}^{\text{($ 2 $)}}_{\lambda = 0, \, s s'} \left( P \right)
      \right],
      \label{eq:fep} \\
    & \partial_\lambda
      \tilde{G}^{\text{($ m $)}}_{\lambda, \, s_1 \ldots s_m} \left( P_1, \ldots, P_{m - 1} \right)
      \notag \\
    & =
      \sum_s
      \tilde{G}^{\text{($ m + 1 $)}}_{\lambda, \, s s_1 \ldots s_m} \left( 0, P_1, \ldots, P_{m - 1} \right)
      \partial_{\lambda}
      J_{\text{sup}, \, \lambda, \, s} \left( \rhom_{\uparrow}, \rhom_{\downarrow} \right)
      \notag \\
    & \,
      \quad
      -
      \frac{1}{2}
      \int_{\bm{p}}
      \tilde{U} \left(\bm{p}\right)
      \left(
      \int_{P^0}
      e^{i P^0 \epsilon'}
      \sum_{s, \, s'}
      \tilde{G}^{\text{($ m + 2 $)}}_{\lambda, \, s s' s_1 \ldots s_m} \left( P, -P, P_1, \ldots, P_{m - 1} \right)
      -
      \sum_s
      \tilde{G}^{\text{($ m + 1 $)}}_{\lambda, \, s s_1 \ldots s_m} \left( 0, P_1, \ldots, P_{m - 1} \right)
      \right)
      \notag \\
    & \,
      \quad
      -
      \frac{1}{2}
      \sum_{k = 1}^{m-1}
      \frac{1}{k! \, \left( m - k\right)!}
      \sum_{\sigma \in S_m}
      \tilde{U} \left( \sum_{i = 1}^k \bm{p}_{\sigma \left( i \right)} \right)
      \sum_s
      \tilde{G}^{\text{($ k + 1 $)}}_{\lambda, \, s s_{\sigma \left( 1 \right)} \ldots s_{\sigma \left( k \right)}}
      \left(
      - \sum_{i = 1}^k
      P_{\sigma \left( i \right)}, P_{\sigma \left( 1 \right)}, \ldots, P_{\sigma \left( k - 1 \right)}
      \right)
      \notag \\
    & \,
      \qquad
      \times
      \sum_{s'}
      \tilde{G}^{\text{($ m - k + 1 $)}}_{\lambda, \, s' s_{\sigma \left( k + 1 \right)} \ldots s_{\sigma \left( m \right)}}
      \left(
      - \sum_{i = 1}^k
      P_{\sigma \left( i \right)}, P_{\sigma \left( k + 1 \right)}, \ldots, P_{\sigma \left( m - 1 \right)}
      \right).
      \label{eq:fgm}
  \end{align}
  Here, we have introduced
  the four vector $ P = \left( P^0, \bm{p} \right) $ with the imaginary frequency $ P^0 $
  and the spatial momentum $ \bm{p} $
  and $ \int_P = \int dP^0 / \left( 2 \pi \right) \int d \bm{p} / \left( 2 \pi \right)^D $.
  The Fourier components
  $ \tilde{G}^{\text{($ m $)}}_{\lambda, \, s_1 \ldots s_m} \left( P_1, \ldots, P_{m - 1} \right) $
  and $ \tilde{U} \left( \bm{p} \right) $
  are defined as follows:
  \begin{align}
    \left( 2 \pi \right)^4
    \delta^4
    \left(
    \sum_{i = 1}^m
    P_i
    \right)
    \tilde{G}^{\text{($ m $)}}_{\lambda, \, s_1 \ldots s_m} \left( P_1, \ldots, P_{m - 1} \right)
    & =
      \int_{X_1, \, \ldots, \, X_m}
      e^{i \sum_{i = 1}^m P_i \cdot X_i}
      G^{\text{($ m $)}}_{\lambda, \, s_1 \ldots s_m} \left( X_1, \ldots, X_m \right),
      \notag \\
    \tilde{U}
    \left( \bm{p} \right)
    =
    \int_X
    e^{i P \cdot X}
    U_{\text{2b}} \left( X \right)
    & =
      \begin{cases}
        2 \pi / \left| \bm{p} \right|
        & \text{($ D = 2 $)}, \\
        4 \pi / \left| \bm{p} \right|^2
        & \text{($ D = 3 $)}.
      \end{cases}
          \label{eq:up}
  \end{align}
  In Eq.~\eqref{eq:fgm},
  one of $ P_{\sigma \left( 1 \right)} $, \ldots, $ P_{\sigma \left( m \right)} $
  becomes $ P_m $, which stands for $ P_m = - \sum_{i = 1}^{m - 1} P_i $.
  \par
  In this paper, we consider the vertex expansion up to the second order.
  The first and second order of Eq.~\eqref{eq:fgm} read
  \begin{align}
    0
    = & \,
        \sum_s
        \tilde{G}^{\text{($ 2 $)}}_{\lambda, \, s s_1} \left( 0 \right)
        \partial_{\lambda}
        J_{\text{sup}, \, \lambda, \, s} \left( \rhom_{\uparrow}, \rhom_{\downarrow} \right)
        -
        \frac{1}{2}
        \int_{\bm{p}}
        \tilde{U} \left( \bm{p} \right)
        \left(
        \int_{P^0}
        e^{i P^0 \epsilon'}
        \sum_{s, \, s'}
        \tilde{G}^{\text{($ 3 $)}}_{\lambda, \, s s' s_1} \left( P, -P \right)
        -
        \sum_s
        \tilde{G}^{\text{($ 2 $)}}_{\lambda, \, s s_1} \left( 0 \right)
        \right), \\
    \partial_{\lambda}
    \tilde{G}^{\text{($ 2 $)}}_{\lambda, \, s_1 s_2} \left( P_1 \right)
    = & \,
        \sum_s
        \tilde{G}^{\text{($ 3 $)}}_{\lambda, \, s s_1 s_2} \left( 0, P_1 \right)
        \partial_{\lambda}
        J_{\text{sup}, \, \lambda, \, s} \left( \rhom_{\uparrow}, \rhom_{\downarrow} \right)
        -
        \tilde{U} \left( \bm{p}_1 \right)
        \sum_{s, \, s'}
        \tilde{G}^{\text{($ 2 $)}}_{\lambda, \, s s_1} \left( -P_1 \right)
        \tilde{G}^{\text{($ 2 $)}}_{\lambda, \, s' s_2} \left( P_1 \right)
        \notag \\
      & \,
        -
        \frac{1}{2}
        \int_{\bm{p}}
        \tilde{U} \left( \bm{p} \right)
        \left(
        \int_{P^0}
        e^{i P^0 \epsilon'}
        \sum_{s, \, s'}
        \tilde{G}^{\text{($ 4 $)}}_{\lambda, \, s s' s_1 s_2} \left( P, -P, P_1 \right)
        -
        \sum_s
        \tilde{G}^{\text{($ 3 $)}}_{\lambda, \, s s_1 s_2} \left( 0, P_1 \right)
        \right).
  \end{align}
  By canceling
  $ \partial_{\lambda}
  J_{\text{sup}, \, \lambda, \, s} \left( \rhom_{\uparrow}, \rhom_{\downarrow} \right) $
  in these equations, we have
  \begin{align}
    \partial_\lambda
    \tilde{G}^{\text{($ 2 $)}}_{\lambda, \, s_1 s_2} \left( P_1 \right)
    = & \,
        -
        \tilde{U} \left( \bm{p}_1 \right)
        \sum_{s, \, s'}
        \tilde{G}^{\text{($ 2 $)}}_{\lambda, \, s s_1} \left( -P_1 \right)
        \tilde{G}^{\text{($ 2 $)}}_{\lambda, \, s' s_2} \left( P_1 \right)
        +
        C_{\lambda, \, s_1 s_2} \left( P_1 \right),
        \label{eq:g2spinflow} \\
    C_{\lambda, \, s_1 s_2} \left( P_1 \right)
    = & \,
        -
        \frac{1}{2}
        \int_{\bm{p}}
        \tilde{U} \left( \bm{p} \right)
        \int_{P^0}
        e^{i P^0 \epsilon'}
        \notag \\
      & \,
        \times
        \sum_{s, \, s'}
        \left(
        \tilde{G}^{\text{($ 4 $)}}_{\lambda, \, s s' s_1 s_2} \left( P, -P, P_1 \right)
        -
        \sum_{t, \, t'}
        \tilde{G}^{\text{($ 3 $)}}_{\lambda, \, t s_1 s_2} \left( 0, P_1 \right)
        \left[
        \tilde{G}^{\text{($ 2 $)}}_{\lambda} \left( 0 \right)
        \right]_{t t'}^{-1}
        \tilde{G}^{\text{($ 3 $)}}_{\lambda, \, s s' t'} \left( P, -P \right)
        \right),
        \label{eq:cdef}
  \end{align}
\end{widetext}
where
$ \left[
  \tilde{G}^{\text{($ 2 $)}}_{\lambda} \left( 0 \right)
\right]_{t t'}^{-1} $
is the inverse of
$ \tilde{G}^{\text{($ 2 $)}}_{\lambda, \, t t'} \left( 0 \right) $
with respect to the spin indices $ t $ and $ t' $.
As shown in Eq.~\eqref{eq:cdef}, $ C_{\lambda, \, s_1 s_2} \left( P_1 \right) $
is composed of higher-order correlation functions.
\par
As in our previous works~\cite{yok19,yok21},
we ignore the $ \lambda $ dependence of $ C_{\lambda, \, s_1 s_2} \left( P_1 \right) $ as
\begin{equation}
  \label{eq:capprox}
  C_{\lambda, \, s_1 s_2} \left( P_1 \right) \approx C_{\lambda = 0, \, s_1 s_2} \left( P_1 \right).
\end{equation}
%$ C_{\lambda, \, s_1 s_2} \left( P_1 \right) \approx C_{\lambda = 0, \, s_1 s_2} \left( P_1 \right)$.
Applying this approximation and summing up the spin indices,
Eqs.~\eqref{eq:fep} and \eqref{eq:g2spinflow} are rewritten as follows:
\begin{align}
  \partial_{\lambda}
  \epsilon_{\text{c}, \, \lambda} \left( \rhom_{\uparrow}, \rhom_{\downarrow} \right)
  = & \,
      \frac{1}{2 \rhom}
      \int_P
      \tilde{U} \left( \bm{p} \right)
      e^{i P^0 \epsilon'}
      \notag \\
    & \,
      \times
      \left[
      \tilde{G}^{\text{($ 2 $)}}_{\lambda} \left( P \right)
      -
      \tilde{G}^{\text{($ 2 $)}}_{\lambda = 0} \left( P \right)
      \right],
      \label{eq:ecflowfinal} \\
  \partial_{\lambda}
  \tilde{G}^{\text{($ 2 $)}}_{\lambda} \left( P_1 \right)
  \approx & \,
            -
            \tilde{U} \left( \bm{p}_1 \right)
            \left[
            \tilde{G}^{\text{($ 2 $)}}_{\lambda} \left( P_1 \right)
            \right]^2
            +
            C_{\lambda=0} \left( P_1 \right),
            \label{eq:g2flowfinal}
\end{align}
where we have introduced the total density correlation functions:
\begin{equation}
  \label{eq:totcol}
  \tilde{G}^{\text{($ m $)}}_{\lambda} \left( P_1, \ldots, P_{m - 1} \right)
  =
  \sum_{s_1, \, \ldots, \, s_m}
  \tilde{G}^{\text{($ m $)}}_{\lambda, \, s_1 \ldots s_m} \left( P_1, \ldots, P_{m - 1} \right),
\end{equation}
and
\begin{equation}
  \label{eq:ccss}
  C_{\lambda=0} \left( P_1 \right)
  =
  \sum_{s, \, s'}
  C_{\lambda=0, s s'} \left( P_1 \right).
\end{equation}
The solution of Eq.~\eqref{eq:g2flowfinal}, which has the form of Riccati equation,
can be obtained analytically with respect to $ \lambda $.
With this solution, the $ \lambda $ integral in Eq.~\eqref{eq:ecflowfinal}
is performed analytically.
The solutions read
\begin{align}
  & \tilde{G}^{\text{($ 2 $)}}_{\lambda = 1} \left( P \right)
    =
    \tilde{G}^{\text{($ 2 $)}}_{\lambda = 0} \left( P \right)
    \frac{1 + \left( B_P / A_P \right) \tanh B_P}{1 + \left( A_P / B_P \right) \tanh B_P},
    \label{eq:gsol} \\
  & \epsilon_{\text{c}, \, \lambda = 1} \left( \rhom_{\uparrow}, \rhom_{\downarrow} \right)
    \notag \\
  & =
    \frac{1}{2 \rhom}
    \int_P
    \left[
    \ln
    \left(
    \cosh B_P
    +
    \frac{A_P}{B_P} \sinh B_P
    \right)
    -
    A_P
    \right],
    \label{eq:ecsol}
\end{align}
where
\begin{subequations}
  \begin{align}
    A_P
    & =
      \tilde{U} \left( \bm{p} \right)
      \tilde{G}^{\text{($ 2 $)}}_{\lambda = 0} \left( P \right),
      \label{eq:ap} \\
    B_P
    & =
      \sqrt{
      \tilde{U} \left( \bm{p} \right)
      C_{\lambda = 0} \left( P \right)}.
      \label{eq:bp}
  \end{align}
\end{subequations}
\par
The quantity $ C_{\lambda=0} \left( P_1 \right) $
is given by the density correlation functions
in the non-interacting case
$ \tilde{G}^{\text{($ m $)}}_{\lambda = 0, \, s_1 \ldots s_m} \left( P_1, \ldots, P_{m - 1} \right) $:
\begin{align}
  & \tilde{G}^{\text{($ m $)}}_{\lambda = 0, \, s_1 \ldots s_m} \left( P_1, \ldots, P_{m - 1} \right)
    \notag \\
  & =
    -
    \sum_{\sigma \in S_{m - 1}}
    \int_{P'}
    \prod_{k = 0}^{m - 1}
    \tilde{G}_{\text{F}, s_{\sigma \left( k \right)} s_{\sigma \left( k + 1 \right)}}
    \left( \sum_{i = 1}^k P_{\sigma \left( i \right)} + P' \right),
    \label{eq:gm}
\end{align}
where
$ \tilde{G}_{\text{F}, s s'} \left( P \right) $
is the propagator of the free fermion defined by
\begin{align}
  \tilde{G}_{\text{F}, s s'} \left( P = \left( \omega, \bm{p} \right) \right)
  & =
    \delta_{s s'}
    \tilde{G}_{\text{F}} \left( p_{\text{F}, s}; P = \left( \omega, \bm{p} \right) \right)
    \notag \\
  & =
    \delta_{s s'}
    \frac{e^{i \omega \epsilon}}{i \omega - \xi_s \left( \bm{p} \right)}
    \label{eq:fermiprop}
\end{align}
with $ \xi_s \left( \bm{p} \right) = \bm{p}^2 / 2 - p_{\text{F}, s}^2 / 2 $
and
the Fermi momentum $ p_{\text{F}, s} = 2 \pi \left( \rhom_s / V_D \right)^{1/D} $.
In Eq.~\eqref{eq:gm}, $ \sigma \left( 0 \right) $ is 
defined by $ \sigma \left( 0 \right) = \sigma \left( m \right) $.
For efficient numerical calculation of $ C_{\lambda = 0} \left( P_1 \right) $,
a technique in Refs.~\cite{yok19,yok21} can be used:
It is convenient to describe $ C_{\lambda = 0} \left( P_1 \right) $
in terms of the total density correlation function
$ \tilde{G}^{\text{($ m $)}}_{\lambda = 0} \left( p_{\text{F}}; P_1, \ldots, P_{m - 1} \right) $
for spin unpolarized systems with the Fermi momentum $ p_{\text{F}} $,
which is defined by
\begin{align}
  & \tilde{G}^{\text{($ m $)}}_{\lambda = 0} \left( p_{\text{F}}; P_1, \ldots, P_{m - 1} \right)
    \notag \\
  & =
    -
    2
    \sum_{\sigma \in S_{m - 1}}
    \int_{P'}
    \prod_{k = 0}^{m - 1}
    \tilde{G}_{\text{F}} \left( p_{\text{F}}; \sum_{i = 1}^k P_{\sigma \left( i \right)} + P' \right),
\end{align}
and related to Eq.~\eqref{eq:gm} as
\begin{align}
  & \tilde{G}^{\text{($ m $)}}_{\lambda = 0, \, s_1 \ldots s_m} \left( P_1, \ldots, P_{m - 1} \right)
    \notag \\
  & =
    \frac{1}{2}
    \left(
    \prod_{k = 1}^{m - 1}
    \delta_{s_k s_{k + 1}}
    \right)
    \tilde{G}^{\text{($ m $)}}_{\lambda = 0} \left( p_{\text{F}, s_1}; P_1, \ldots, P_{m - 1} \right).
    \label{eq:gmgmp}
\end{align}
Then, $ C_{\lambda = 0, \, s_1 s_2} \left( P_1 \right) $ is written as follows:
\begin{equation}
  \label{eq:ccp}
  C_{\lambda = 0, \, s_1 s_2} \left( P_1 \right)
  =
  \frac{\delta_{s_1 s_2}}{2}
  C_{\lambda = 0} \left( p_{\text{F}, s_1}; P_1 \right),
\end{equation}
where
\begin{align}
  & C_{\lambda = 0} \left( p_{\text{F}}; P_1 \right)
    \notag \\
  & =
    -
    \frac{1}{2}
    \int_{\bm{p}}
    \tilde{U} \left( \bm{p} \right)
    \int_{P^0}
    e^{i P^0 \epsilon'}
    \left[
    \tilde{G}^{\text{($ 4 $)}}_{\lambda = 0} \left( p_{\text{F}}; P, -P, P_1 \right)
    \right.
    \notag \\
  & \quad
    \left.
    -
    \frac{
    \tilde{G}^{\text{($ 3 $)}}_{\lambda = 0} \left( p_{\text{F}}; P, -P \right)
    \tilde{G}^{\text{($ 3 $)}}_{\lambda = 0} \left( p_{\text{F}}; P_1, -P_1 \right)}
    {\tilde{G}^{\text{($ 2 $)}}_{\lambda = 0} \left( p_{\text{F}}; 0 \right)}
    \right].
    \label{eq:c0}
\end{align}
As shown in Refs.~\cite{yok19,yok21},
the momentum integrals in Eq.~\eqref{eq:c0} can be reduced to double integrals,
which reduces the computational time and enables us to calculate
$ \epsilon_{\text{c}, \, \lambda = 1} $
in a few minutes even on a laptop computer
for each set of $ \left( \rs, \zeta \right) $.
\subsection{Validity of the approximation}
%Dense limit \cor{and finite $\rs$ case}
\label{sec:denselimit}
\par
%\cor{
We discuss the validity of our approximation described by Eq.~\eqref{eq:capprox}.
First, we show that the resultant
$ \epsilon_{\text{c}, \, \lambda = 1} $
given by Eq.~\eqref{eq:ecsol}
%}
%\sout{Equation~\eqref{eq:ecsol}}
reproduces the exact result at the dense limit given by the GB resummation.
To show this, we employ the following scaling rules for
$ C_{\lambda = 0, \, s_1 s_2} \left( p_{\text{F}}; P_1 \right) $
and
$ \tilde{G}^{\text{($ 2 $)}}_{\lambda = 0, \,  s_1 s_2} \left( p_{\text{F}}; P_1 \right) $:
\begin{subequations}
  \begin{align}
    \tilde{G}^{\text{($ 2 $)}}_{\lambda = 0, \,  s_1 s_2} \left( p_{\text{F}}; P_1 \right)
    & =
      a^{D - 2}
      \tilde{G}^{\text{($ 2 $)}}_{\lambda = 0, \,  s_1 s_2}
      \left( \frac{p_{\text{F}}}{a}; \left( \frac{P_1^0}{a^2}, \frac{\bm{p}_1}{a} \right) \right), \\
    C_{\lambda=0, s_1 s_2} \left( p_{\text{F}}; P_1 \right)
    & =
      a^{D - 3}
      C_{\lambda=0, s_1 s_2}
      \left( \frac{p_{\text{F}}}{a}; \left( \frac{P_1^0}{a^2}, \frac{\bm{p}_1}{a} \right) \right),
  \end{align}
\end{subequations}
for an arbitrary number $ a $.
By applying these rules to
Eqs.~\eqref{eq:totcol}, \eqref{eq:ccss}, \eqref{eq:gmgmp}, and \eqref{eq:ccp}
with $ a = \rs^{-1} $ and using
$ \rs p_{\text{F}, s} = 2 \pi \left[ \left( 1 + s \zeta \right) / \left(2 V_D^2 \right) \right]^{1/D} $,
we obtain
\begin{subequations}
  \begin{align}
    \tilde{G}^{\text{($ 2 $)}}_{\lambda = 0} \left( P_1 \right)
    & =
      \rs^{2 - D}
      \tilde{\mathcal{G}}^{\text{($ 2 $)}}_{\lambda = 0} \left( \overline{P}_1 \right),
      \label{eq:gscale} \\
    C_{\lambda = 0} \left( P_1 \right)
    & =
      \rs^{3 - D}
      \mathcal{C}_{\lambda = 0} \left( \overline{P}_1 \right).
      \label{eq:cscale}
  \end{align}
\end{subequations}
Here, we have introduced
$ \overline{P}_1 = \left( \rs^2 P_1^0, \rs \bm{p}_1 \right) $ 
and
\begin{subequations}
  \begin{align}
    \tilde{\mathcal{G}}^{\text{($ 2 $)}}_{\lambda = 0} \left( \overline{P}_1 \right)
    & =
      \frac{1}{2}
      \sum_s
      \tilde{G}^{\text{($ 2 $)}}_{\lambda = 0}
      \left( 2 \pi \left( \frac{1 + s \zeta}{2 V_D^2} \right)^{1/D} ; \overline{P}_1 \right), \\
    \mathcal{C}_{\lambda = 0} \left( P_1 \right)
    & =
      \frac{1}{2}
      \sum_s
      C_{\lambda = 0}
      \left( 2 \pi \left( \frac{1 + s \zeta}{2 V_D^2} \right)^{1/D} ; \overline{P}_1 \right),
  \end{align}
\end{subequations}
which are independent of $ \rs $ except for $ \overline{P}_1 $.
Using Eqs.~\eqref{eq:gscale} and \eqref{eq:cscale}, and 
$ \tilde{U} \left( \bm{p} \right) = \rs^{D - 1} \tilde{U} \left( \overline{\bm{p}} \right) $
given by Eq.~\eqref{eq:up},
we obtain the scaling for $ A_P $ and $ B_P $ as follows:
\begin{subequations}
  \begin{align}
    A_P
    & =
      \rs \mathcal{A}_{\overline{P}}, \\
    B_P
    & =
      \rs \mathcal{B}_{\overline{P}},
  \end{align}
\end{subequations}
where we have introduced
\begin{subequations}
  \begin{align}
    \mathcal{A}_{\overline{P}}
    & =
      \tilde{U} \left( \overline{\bm{p}} \right)
      \tilde{\mathcal{G}}^{\text{($ 2 $)}}_{\lambda = 0} \left( \overline{P} \right), \\
    \mathcal{B}_{\overline{P}}
    & =
      \sqrt{\tilde{U} \left( \overline{\bm{p}} \right)
      \mathcal{C}_{\lambda = 0} \left( \overline{P} \right)},
  \end{align}
\end{subequations}
which are independent of $ \rs $ except for $ \overline{P} $.
Changing the integral variable $ P $ to $ \overline{P} $ in Eq.~\eqref{eq:ecsol}
and using $ \rs = \left( V_D \rhom \right)^{-1/D} $, we have
\begin{align}
  & \epsilon_{\text{c}, \, \lambda=1} \left( \rhom_{\uparrow}, \rhom_{\downarrow} \right)
    \notag \\
  & =
    \frac{V_D^{1/D}}{2 \rs^2}
    \int_{\overline{P}}
    \left(
    \ln \left[
    \cosh \left( \rs \mathcal{B}_{\overline{P}} \right)
    +
    \frac{\mathcal{A}_{\overline{P}}}{\mathcal{B}_{\overline{P}}}
    \sinh \left( \rs \mathcal{B}_{\overline{P}} \right)
    \right]
    -
    \rs \mathcal{A}_{\overline{P}}
    \right).
\end{align}
By expanding this equation with respect to $ \rs $, we obtain
\begin{align}
  & \epsilon_{\text{c}, \, \lambda=1} \left( \rhom_{\uparrow}, \rhom_{\downarrow} \right)
    \notag \\
  & =
    \frac{V_D^{1/D}}{2 \rs^2}
    \int_{\overline{P}}
    \left(
    \ln \left( 1 + \rs \mathcal{A}_{\overline{P}} \right)
    -
    \rs \mathcal{A}_{\overline{P}}
    +
    \frac{\left( \rs \mathcal{B}_{\overline{P}} \right)^2}{2}
    \right)
    +
    \mathcal{O} \left( \rs \right)
    \notag \\
  & =
    \frac{1}{2 \rhom}
    \int_P
    \left[
    \ln \left(
    1
    +
    \tilde{U} \left( \bm{p} \right)
    \tilde{G}^{\text{($ 2 $)}}_{\lambda = 0} \left( P \right)
    \right)
    -
    \tilde{U} \left( \bm{p} \right)
    \tilde{G}^{\text{($ 2 $)}}_{\lambda = 0} \left( P \right)
    \right]
    \notag \\
  & \quad
    +
    \frac{1}{4 \rhom}
    \int_P
    \tilde{U} \left( \bm{p} \right)
    C_{\lambda = 0} \left( P \right)
    +
    \mathcal{O} \left( \rs \right).
    \label{eq:ecexpand}
\end{align}
The first term of the last line is the contribution from the random phase approximation.
By performing the frequency integrals, the second term is evaluated as follows:
\begin{align}
  & \frac{1}{4 \rhom}
    \int_P
    \tilde{U} \left( \bm{p} \right)
    C_{\lambda = 0} \left( P \right)
    \notag \\
  & =
    \frac{1}{2 \rhom}
    \sum_s
    \iiint_{\bm{p}, \, \bm{p}', \, \bm{p}''}
    \frac{
    \tilde{U} \left( \bm{p} \right)
    \tilde{U} \left( \bm{p} + \bm{p}' + \bm{p}'' \right)}
    {\bm{p} \cdot \left( \bm{p} + \bm{p}' + \bm{p}'' \right)}
    \notag \\
  & \quad
    \times
    \theta \left( - \xi_s \left( \bm{p}' \right) \right)
    \left[ 1 - \theta \left( - \xi_s \left( \bm{p} + \bm{p}' \right) \right) \right]
    \notag \\
  & \quad
    \times
    \theta \left( - \xi_s \left( \bm{p}'' \right) \right)
    \left[ 1 - \theta \left( - \xi_s \left( \bm{p} + \bm{p}'' \right) \right) \right],
\end{align}
which is identical to the expression for the second-order exchange term~\cite{gel57}.
In summary, Eq.~\eqref{eq:ecexpand} is identical to the expression for the GB resummation.
\par
%\cor{
Although our approximation reproduces
the exact behavior at $ \rs \to 0 $ as shown above,
%is reproduced,
%the ignorance of the flow of the higher-order correlation functions
%represented by Eq.~\eqref{eq:capprox}
%affects the result at finite $\rs$.
%The fact that
the flow of the higher-order correlation functions
becomes important for the accurate calculation
as $ \rs $ increases.
This can be seen from Eq.~\eqref{eq:fgm} 
at arbitrary order roughly:
For simplicity, we consider the case of $ \lambda = 0 $.
The following scaling holds:
\begin{align}
  & \tilde{G}^{\text{($ m $)}}_{\lambda = 0, \, s_1 \ldots s_m}
    \left( P_1, \ldots, P_{m - 1} \right)
    \notag \\
  & =
    \rs^{2m-2-D}
    \tilde{\mathcal{G}}^{\text{($ m $)}}_{\lambda = 0, \, s_1 \ldots s_m}
    \left( \overline{P}_1, \ldots, \overline{P}_{m - 1} \right),
\end{align}
where 
$ \tilde{\mathcal{G}}^{\text{($ m $)}}_{\lambda = 0, \, s_1 \ldots s_m}
\left( \overline{P}_1, \ldots, \overline{P}_{m - 1} \right) $
is a function independent of $ \rs $ given by
\begin{align}
  & \tilde{\mathcal{G}}^{\text{($ m $)}}_{\lambda = 0, \, s_1 \ldots s_m}
    \left( \overline{P}_1, \ldots, \overline{P}_{m - 1} \right)
    \notag \\
  & = 
    \frac{1}{2}
    \left(
    \prod_{k = 1}^{m - 1}
    \delta_{s_k s_{k + 1}}
    \right)
    \notag \\
  & \qquad
    \times
    \tilde{G}^{\text{($ m $)}}_{\lambda = 0}
    \left( 
    2 \pi \left( \frac{1 + s_1 \zeta}{2 V_D^2} \right)^{1/D};
    \overline{P}_1, \ldots, \overline{P}_{m - 1}
    \right).
\end{align}
By use of this scaling, we find that
the flow represented by Eq.~\eqref{eq:fgm} 
at $ \lambda = 0 $ behaves as
\begin{equation}
  \left.
  \frac{
  \partial_{\lambda}
  \tilde{G}^{\text{($ m $)}}_{\lambda, \, s_1 \ldots s_m} \left( P_1, \ldots, P_{m - 1} \right)}
  {\tilde{G}^{\text{($ m $)}}_{\lambda, \, s_1 \ldots s_m} \left( P_1, \ldots, P_{m - 1} \right)}
  \right|_{\lambda=0}
  \sim
  \rs,
\end{equation}
which shows that
$ \tilde{G}^{\text{($ m $)}}_{\lambda, \, s_1 \ldots s_m} \left( P_1, \ldots, P_{m - 1} \right) $
rapidly evolves as $ \rs $ increases.
This result indicates that the evolution of
$ \tilde{G}^{\text{($ m \geq 3 $)}}_{\lambda, \, s_1 \ldots s_m} \left( P_1, \ldots, P_{m - 1} \right) $,
which is ignored in our approximation given by Eq.~\eqref{eq:capprox},
becomes important for the accuracy at large $ \rs $.
%}
% 
\section{Numerical results}
\label{sec:result}
\begin{figure}[tb]
  \centering
  \includegraphics[width=\columnwidth]{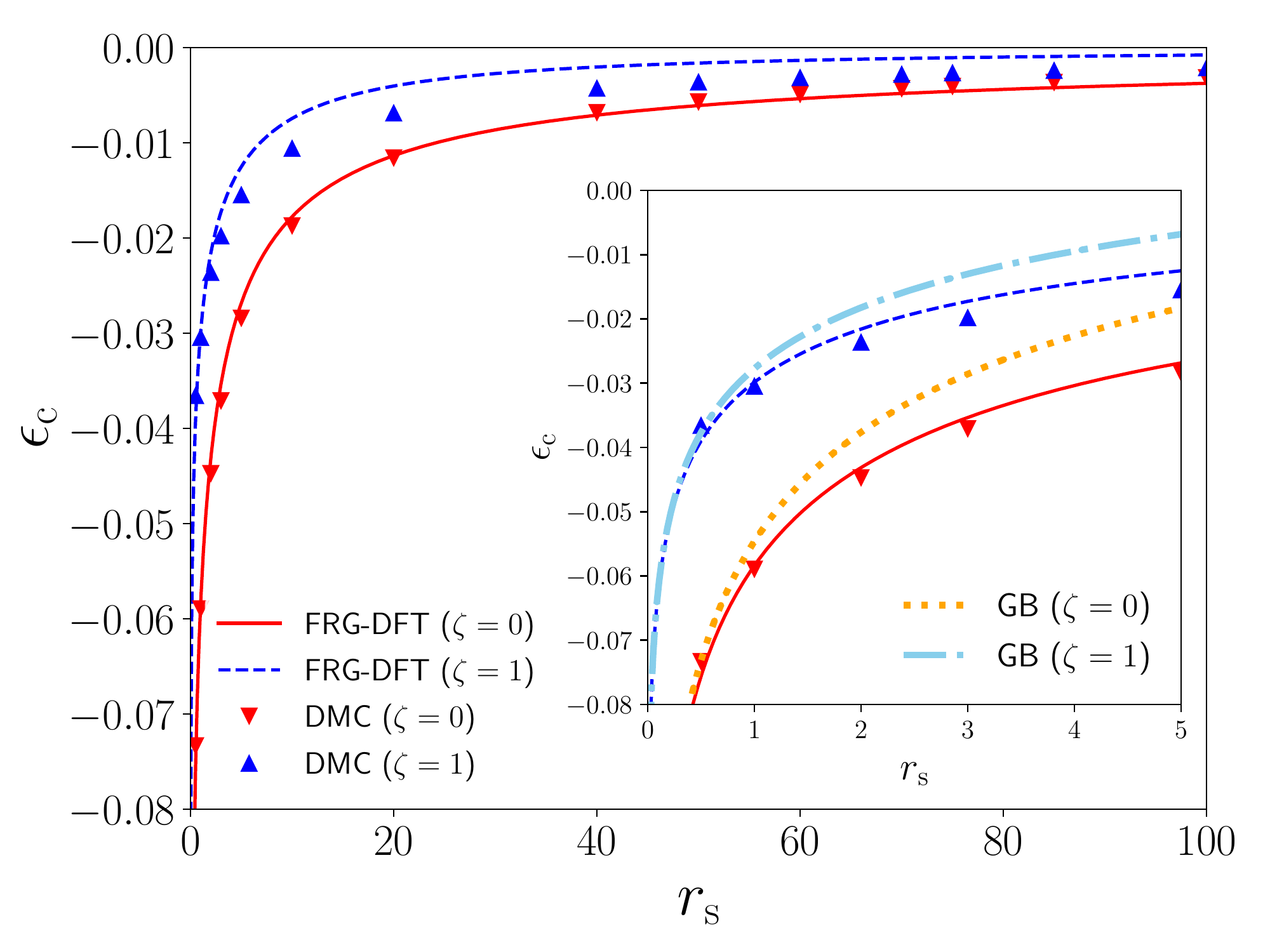}
  \caption{Correlation energy per particle $ \ec $
    for three-dimensional ($ D = 3 $) homogeneous electron gas
    in the cases of $ \zeta = 0 $ (paramagnetic) and $ 1 $ (ferromagnetic)
    calculated by using the FRG-DFT (this work) and DMC shown as functions of Wigner-Seitz radius $ \rs $.
    The data of DMC calculations are obtained from the values in Table IV in Ref.~\cite{loo16},
    which summarizes the results in Refs.~\cite{zon02,spi13}.
    The inset is the enlarged view in $ 0 \leq \rs \leq 5 $,
    where the results by the GB resummation are also shown.
    }
  \label{fig:three_zeta01}
\end{figure}
\begin{table*}[tb]
  \centering
  \caption{Absolute difference 
  $ \Delta \ec = \left| \ec^{\text{FRG}} - \ec^{\text{DMC}} \right| $
  and relative difference $ \Delta \ec / \left| \ec^{\text{DMC}} \right| $
  between the FRG-DFT result $ \ec^{\text{FRG}} $ (this work) and
  DMC result $ \ec^{\text{DMC}} $~\cite{loo16} 
  at $ \rs = 1 $, $ 2 $, $ 3 $, $ 5 $, $ 10 $, $ 50 $, and $ 100 \, \mathrm{a.u.} $
  and $ \zeta = 0 $ and $ 1 $.}
  \label{tab:corr}
  \begin{ruledtabular}
    \begin{tabular}{ccddddddd}
      & \multicolumn{1}{c}{$ \rs $ ($ \mathrm{a.u.} $)}
      & \multicolumn{1}{c}{$ 1 $}
      & \multicolumn{1}{c}{$ 2 $}
      & \multicolumn{1}{c}{$ 3 $}
      & \multicolumn{1}{c}{$ 5 $}
      & \multicolumn{1}{c}{$ 10 $}
      & \multicolumn{1}{c}{$ 50 $}
      & \multicolumn{1}{c}{$ 100 $} \\
      \hline
      $ \zeta = 0 $
      &
      $ \Delta \ec \times 10^4 $
      & 6.0 & 16.4 & 17.9 & 16.1 &  9.9 &  3.6 &  5.5 \\
      &
      $ \Delta \ec / |\ec^{\rm DMC}| $ ($\%$)
      & 1.0 &  3.7 & 4.8 & 5.7 &  5.3 & 6.2 &  17.0 \\
      $ \zeta = 1 $
      &
      $ \Delta \ec \times 10^4 $
      & 5.6 & 20.1 & 24.9 & 29.2 & 31.1 & 19.6 & 13.0 \\
      &
      $ \Delta \ec / \left| \ec^{\text{DMC}} \right| $ ($\%$)
      & 1.8 & 8.5  & 12.6 & 18.9 &  29.5 & 54.8 & 62.6 \\
    \end{tabular}
  \end{ruledtabular}
\end{table*}
\par
In this section, we show the numerical results for HEG.
Figure~\ref{fig:three_zeta01} shows the FRG-DFT results
of $ \epsilon_{\text{c}, \, \lambda = 1} $ in three dimensions
in the paramagnetic ($ \zeta = 0 $) and ferromagnetic ($ \zeta = 1 $) states, together with
the results by the diffusion Monte Carlo (DMC) simulation
and the GB resummation,
as functions of the Wigner-Seitz radius 
$ \rs $.~\footnote{
  There is a tiny difference
  between the result of FRG-DFT at $ \zeta = 0 $ in this work
  and that in Fig.~1 in Ref.~\cite{yok21}.
  We find that a coefficient is underestimated
  in the numerical code to obtain the latter one.
  The present result is based on a corrected code.}
The DMC results are obtained by subtracting
the kinetic and exchange energies per particle
from the total energy per particle
given in Table IV in Ref.~\cite{loo16},
which summarizes the results in Refs.~\cite{zon02,spi13}.
%\sout{The FRG-DFT result in the case of $ \zeta = 0 $ is the same as that shown in Ref.~\cite{yok21}.}
For both cases of $ \zeta = 0 $ and $ 1 $,
the FRG-DFT reproduces the results by the GB resummation and
the discrepancies between the FRG-DFT and DMC results
decrease as $ \rs $ becomes close to $ 0 $,
which is also indicated by the relative differences shown in
Table~\ref{tab:corr}.
%\sout{reflects the property that the FRG-DFT
%reproduce the exact behavior given by the GB resummation at $ \rs \to 0 $.}
On the other hand, the increase of
the relative differences at larger $ \rs $
is due to the ignorance of
the flow of the higher-order correlation functions
as discussed in Sec.~\ref{sec:denselimit}.
In the case of $ \zeta = 1 $,
FRG-DFT overestimates $ \ec $ compared to DMC.
This is a natural result since the truncation up to the second order
breaks the Pauli-blocking condition \cite{kem17a,yok18},
which allows two electrons with the same spin
to get closer to each other and increases the energy.
At $ \zeta = 0 $ and $ \rs \gtrsim 40 \, \mathrm{a.u.} $,
FRG-DFT underestimates $ \ec $
with smaller deviation compared to the case of $ \zeta = 1 $
as indicated in the absolute and relative differences shown in
Table~\ref{tab:corr}.
This suggests that the correlation
between two electrons with the different spins
is underestimated and compensates
the overestimation coming from the
correlation between those with the same spin.
%\sout{The differences in the case of $ \zeta = 1 $ are greater than those in the case of $ \zeta = 0$
%as indicated in Table~\ref{tab:corr}.
%This may be because the case of $ \zeta = 1 $,
%where all the fermions have the same component,
%is more sensitive to the breaking of the Pauli blocking effect,
%which is caused by the truncation up to the second order
%as discussed in Refs.~[18,19].}
% 
\begin{figure}[tb]
  \centering
  \includegraphics[width=\columnwidth]{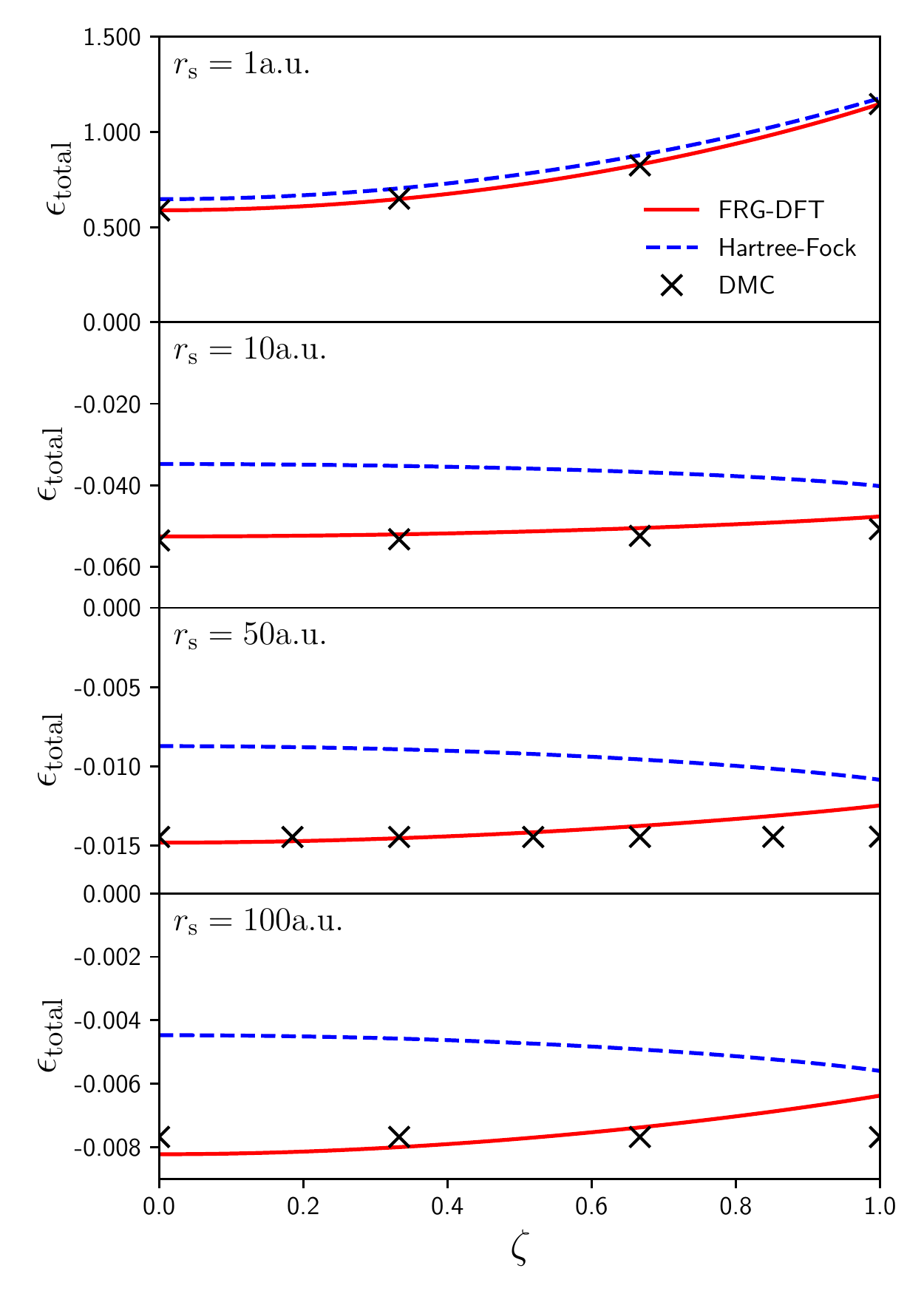}
  \caption{Total energy per particle $ \epsilon_{\text{total}} $
    at $ \rs = 1 $, $ 10 $, $ 50 $, and $ 100 \, \mathrm{a.u.} $
    calculated by using the FRG-DFT
    shown as functions of $ \zeta $.
    The results from the Hartree--Fock approximation and DMC
    are also shown.
    The data of DMC calculations are taken from Table IV in Ref.~\cite{loo16},
    which summarizes the results in Refs.~\cite{zon02,spi13}.}
  \label{fig:three_zetadep}
\end{figure}
\par
For the purpose to discuss stability of phases,
we show the FRG-DFT results of
the total energy per particle $ \epsilon_{\text{total}} $
for arbitrary spin polarization at
$ \rs = 1 $, $ 10 $, $ 50 $, and $ 100 \, \mathrm{a.u.} $ in Fig.~\ref{fig:three_zetadep}.
For comparison, this figure also shows the DMC results and the Hartree--Fock results.
The latter are given by
\begin{equation}
  \epsilon_{\text{total}} \left( \rs, \zeta \right)
  \approx
  \epsilon_{\text{kin}} \left( \rs, \zeta \right)
  +
  \epsilon_{\text{x}, \, \lambda = 1} \left( \rs, \zeta \right),
\end{equation}
where the kinetic term $\epsilon_{\text{kin}}$
and the exchange term $ \epsilon_{\text{x}} $
are given by
\begin{align}
  \epsilon_{\text{kin}} \left( \rs, \zeta \right)
  =
  \frac{3}{10}
  \left( \frac{9\pi}{4} \right)^{2/3}
  \frac{\left( 1 + \zeta \right)^{5/3} + \left( 1 - \zeta \right)^{5/3}}{2 \rs^2},
\end{align}
and Eq.~\eqref{eq:ex}, respectively.
According to the DMC results~\cite{zon02},
a second order phase transition to spin-polarized states
is found at $ \rs = 50 \pm 2 \, \mathrm{a.u.} $ and
the system becomes (partially) spin-polarized states for larger $ \rs $ and unpolarized one for smaller $ \rs $.
In contrast to this,
the unpolarized state is stable
even in $ \rs \gtrsim 50 \, \mathrm{a.u.} $ in the FRG-DFT result.
\par
\begin{figure}[tb]
  \centering
  \includegraphics[width=\columnwidth]{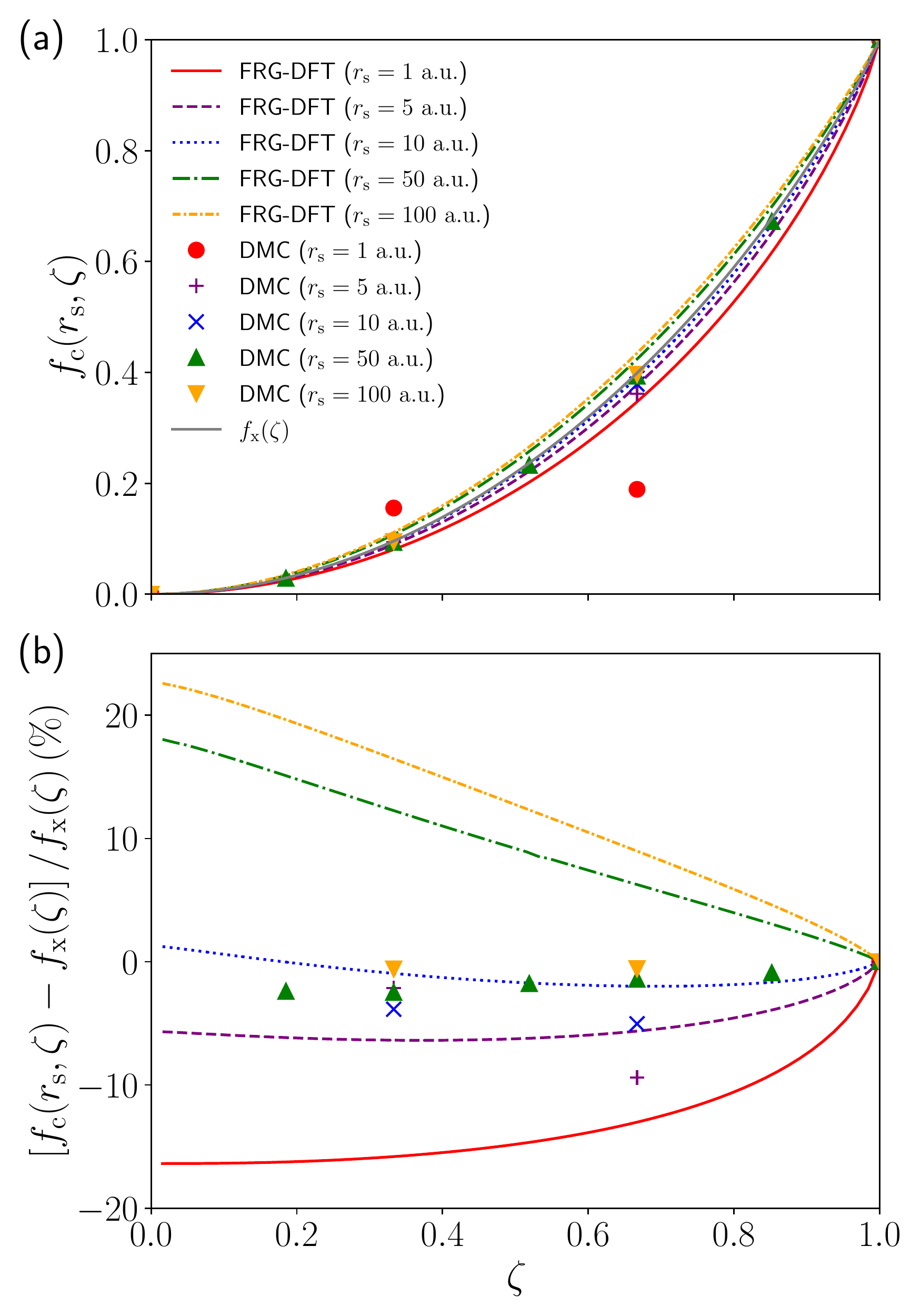}
  \caption{(a) Interpolation function
    $ f_{\text{c}} \left( \rs, \zeta \right) $
    for $ \rs = 1 $, $ 5 $, $ 10 $, $ 50 $, and $ 100 \, \mathrm{a.u.} $
    calculated by using the FRG-DFT as functions of $ \zeta $.
    For comparison, DMC results~\cite{zon02,spi13} and
    $ f_{\text{x}} \left( \zeta \right) $ [Eq.~\eqref{eq:fx}] are also shown.
    (b) Relative deviation of $ f_{\text{c}} \left( \rs, \zeta \right) $
    from $ f_{\text{x}} \left( \zeta \right) $.
    The data at $ \zeta = 0 $, where $ f_{\text{x}} \left( \zeta \right) = 0 $, are excluded.
    The DMC data at $ \rs = 1 \, \mathrm{a.u.} $
    are out of the range of the figure.}
  \label{fig:f_zeta}
\end{figure}

Next, we shall discuss the interpolation function
$ f_{\text{c}} \left( \rs, \zeta \right) $ defined in Eq.~\eqref{eq:fc}.
In Fig.~\ref{fig:f_zeta}(a),
the results for $ \rs = 1 $, $ 5 $, $ 10 $, $ 50 $, and $ 100 \, \mathrm{a.u.} $
calculated by using the FRG-DFT are shown as functions of $ \zeta $.
For comparison, DMC results and
the interpolation function for the exchange part
$ f_{\text{x}} \left( \zeta \right) $ defined in Eq.~\eqref{eq:fx}
are also shown.
Figure \ref{fig:f_zeta}(b) shows
the relative deviation of $ f_{\text{c}} \left( \rs, \zeta \right) $
from $ f_{\text{x}} \left( \zeta \right) $:
$\left[f_{\text{c}} \left( \rs, \zeta \right)-f_{\text{x}} \left( \zeta \right)
\right]/f_{\text{x}} \left( \zeta \right)$.
A conventional approximation for $ f_{\text{c}} \left( \rs, \zeta \right) $ [Eq.~\eqref{eq:fc}]
is $ f_{\text{c}} \left( \rs, \zeta \right) \approx f_{\text{x}} \left( \zeta \right) $~\cite{
  Barth1972J.Phys.C5_1629,
  Martin2004_CambridgeUniversityPress}.
Actually, the DMC results in Fig.~\ref{fig:f_zeta}
suggest that
this approximation is valid for $ \rs \gtrsim 10 \, \mathrm{a.u.} $
However, the FRG-DFT results show stronger $ \rs $ dependence.
Particularly, the deviation of the FRG-DFT results
from $ f_{\text{x}} \left( \zeta \right) $
can be seen in small $ \rs $, where the FRG-DFT is accurate,
as well as in large $ \rs $.
The deviation from $ f_{\text{x}} \left( \zeta \right) $
also can be seen in the DMC results in $ \rs \lesssim 5 \, \mathrm{a.u.} $
\begin{figure}[tb]
  \centering
  \includegraphics[width=\columnwidth]{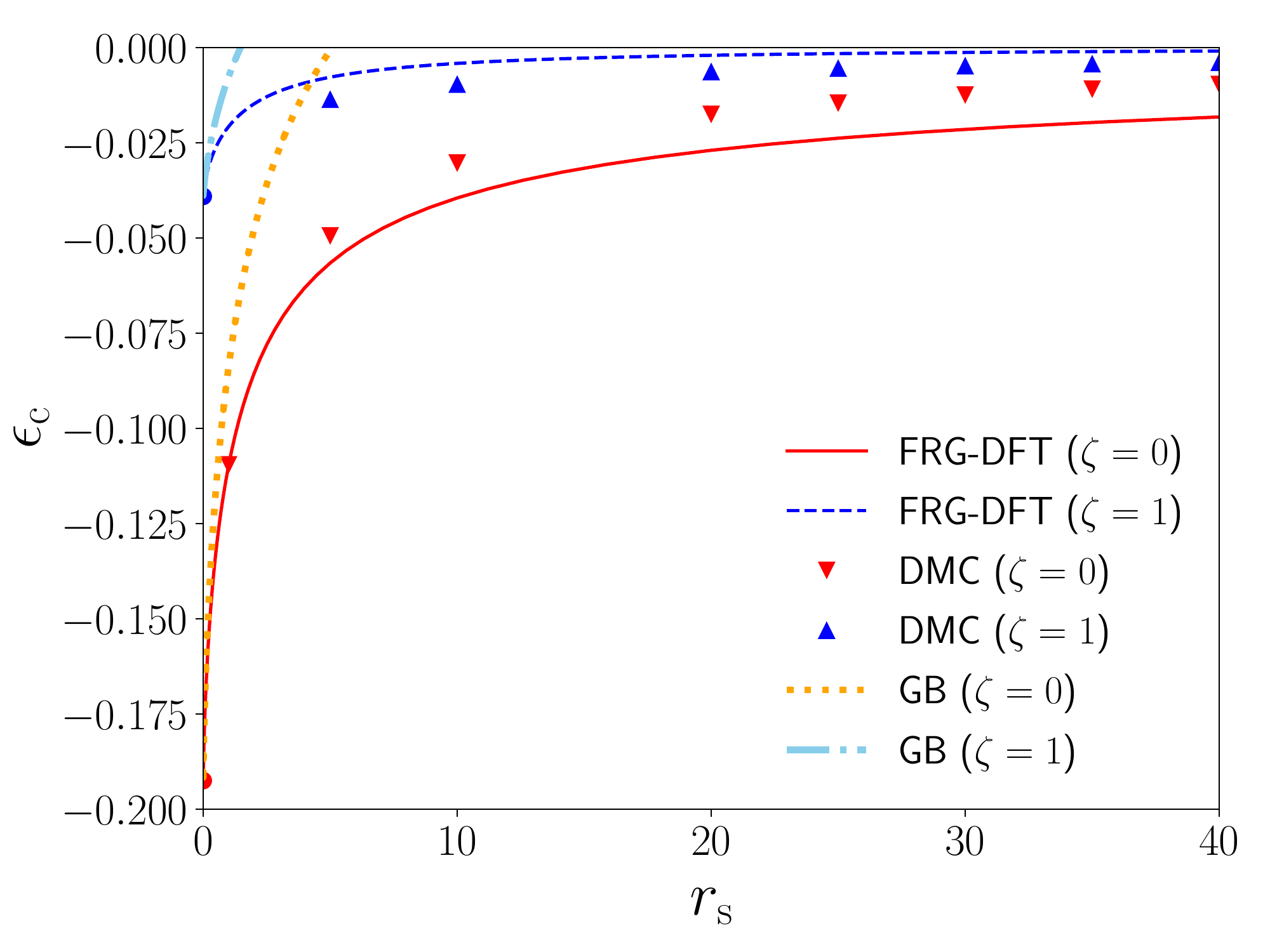}
  \caption{
%  \cor{
  Correlation energy per particle $\ec$
  at $ \zeta = 0 $ and $ 1 $
  given by the FRG-DFT, DMC, and GB resummation in the
  two-dimensional ($ D = 2 $) case.
  The data of DMC calculations are obtained from the values in Table VI in Ref.~\cite{loo16},
  which summarizes the results in Refs.~\cite{kwo93,rap96,dru09}.
The values at $\rs = 0$ given by the FRG-DFT (and the GB resummation), which are finite in contrast to the three-dimensional case, are shown as the red and blue circles for $ \zeta = 0 $ and $1$, respectively.}
  \label{fig:2d_rs}
\end{figure}
\par
Finally, we mention the case of two dimensions.
Figure~\ref{fig:2d_rs} shows the FRG-DFT and DMC results
of $ \rs $ dependence of $ \ec $ at $ \zeta = 0 $ and $ 1 $.
The FRG-DFT result at $ \zeta = 0 $ is the same as that in Ref.~\cite{yok19}.
The DMC results are obtained from the total energies given in Refs.~\cite{kwo93,rap96,dru09},
which are summarized in Table VI in Ref.~\cite{loo16}.
In dense cases, reproducing the exact results at $ \rs = 0 \, \mathrm{a.u.} $ given by the GB resummation,
the FRG-DFT results agree with the DMC results.
On the other hand, the discrepancy between the FRG-DFT and DMC results increases as
the system becomes dilute and the FRG-DFT respectively gives
underestimated and overestimated results
at $ \zeta = 0 $ and $ 1 $ in comparison with DMC.
This behavior of $ \ec $ favors
the paramagnetic phase even if the system is dilute as in the case of three dimensions.
\section{Conclusion}
\label{sec:conc}
\par
We have developed the functional-renormalization-group aided density functional theory (FRG-DFT)
for the description of arbitrary spin-polarized systems
and achieved numerical derivation of the correlation energy per particle
of homogeneous electron gas $ \ec $
with arbitrary density and spin polarization,
which gives the energy density functional in the local spin density approximation.
The hierarchical flow equations for the density correlation functions
have been derived up to arbitrary order based on the FRG-DFT flow equation.
Our numerical calculation has been performed based on the second order truncation
for the hierarchical equations.
Our correlation energy per particle
reproduces the exact behavior at high-density limit
given by the Gell-Mann--Brueckner resummation and agrees with the diffusion Monte Carlo (DMC) results in relatively high-density cases.
On the other hand, the discrepancy between the FRG-DFT and DMC results
becomes significant in the spin-polarized case
in comparison with the spin-unpolarized case as the system becomes dilute.
In contrast to DMC results, the correlation energy given by FRG-DFT stabilizes
the spin-unpolarized state even in dilute cases.
We also discuss the interpolation function
$ f_{\text{c}} \left( \rs, \zeta \right) $,
which characterizes the $ \zeta $ dependence of  $ \ec $.
We find the deviation from 
the interpolation function for the exchange part
at small $ \rs $, where the FRG-DFT gives accurate results.
\par
The growth of the discrepancy in the spin-polarized case
may be attributed to the effect of the Pauli blocking, which is broken in our approximation.
In order to retain the Pauli-blocking effect,
one may introduce a correction factor
to the four-point density correlation function~\cite{kem17a}.
%\cor{
When applying this method, it is required
to solve the flow equation numerically with respect to the evolution parameter $\lambda$.
This is in contrast to the fact that the flow equation can be solved analytically in the approximation in this paper.
%}
%which is not necessary with the approximation
%in this paper since the flow equation can be analytically solved.
The introduction of another approximation scheme
valid even for dilute systems also may change the situation.
In Ref.~\cite{dup21}, a possibility of using small expansion parameters
within some frameworks including the derivative expansion
is discussed.
%\cor{
As for the application of the derivative expansion
in the framework of the functional renormalization group based on density,
there is a work for classical liquids~\cite{lue15}.
%}
\par
A great goal of studies of the FRG-DFT is systematic inclusion of the gradient effect.
In dilute cases, particularly, this is important for the description of the Wigner crystal.
One of the ways to realize this in our framework
may be the use of the derivative expansion.
Methods to describe solid-liquid phase transition
developed for the classical DFT~\cite{ram79} are also expected to give hints for the treatment of the Wigner crystal.
\begin{acknowledgements}
  The authors thank Haozhao Liang for discussions at the early stage of this work.
  T.Y.~was supported by the RIKEN Special Postdoctoral Researchers Program.
  T.N.~was supported by the Grants-in-Aid for JSPS fellows (Grant No.~19J20543).
  Numerical computation in this work was carried out at the Yukawa Institute Computer Facility.
\end{acknowledgements}
\appendix
\begin{widetext}
  \section{Derivation of Eqs.~\eqref{eq:flow0} and \eqref{eq:flown}}
  \label{sec:deriv}
  \par
  In this Appendix, we show the derivation of
  Eqs.~\eqref{eq:flow0} and \eqref{eq:flown}.
  Equation~\eqref{eq:flow0} is derived from Eq.~\eqref{eq:flow_eq}
  and
  \begin{equation}
    \label{eq:gam2g2}
    \Gamma^{\text{($ 2 $)} {-1}}_{\lambda, \, s s'}
    \left[ \rho_{\uparrow}, \rho_{\downarrow} \right] \left( X, X' \right)
    =
    G^{\text{($ 2 $)}}_{\lambda, \, s s'} \left( X, X' \right).
  \end{equation}
  The derivation of this relation is as follows:
  Differentiating Eq.~\eqref{eq:jsup}, we have
  \begin{equation}
    \label{eq:jrw2}
    \sum_t
    \int_Y
    \frac{\delta J_{\text{sup}, \, \lambda, \, t} \left[ \rhou, \rhod \right] \left( Y \right)}
    {\delta \rho_{s'} \left( X' \right)}
    \frac{\delta^2 W_{\lambda}
      \left[
        J_{\text{sup}, \, \lambda, \, {\uparrow}} \left[ \rhou, \rhod \right],
        J_{\text{sup}, \, \lambda, \, {\downarrow}} \left[ \rhou, \rhod \right]
      \right]}
    {\delta J_t \left( Y \right) \, \delta J_s \left( X \right)}
    =
    \delta_{s s'}
    \delta \left( X - X' \right).
  \end{equation}
  By use of Eqs.~\eqref{eq:gamma1} and \eqref{eq:gdef}, this is rewritten as
  \begin{equation}
    \sum_t
    \int_Y
    \frac{\delta^2 \Gamma_{\lambda} \left[ \rhou, \rhod \right]}
    {\delta \rho_{s'} \left( X' \right) \, \delta \rho_t \left( Y \right)}
    G^{\text{($ 2 $)}}_{\lambda, \, t s} \left( Y, X \right)
    =
    \delta_{s s'}
    \delta \left( X - X' \right),
  \end{equation}
  which is equivalent to Eq.~\eqref{eq:gam2g2}.
  \par
  The following relation is useful for the derivation of Eq.~\eqref{eq:flown}:
  \begin{equation}
    \label{eq:gm1}
    \sum_s
    \int_X
    G^{\text{($ 2 $)}}_{\lambda, \, s_{m + 1} s} \left( X_{m + 1}, X \right)
    \frac{\delta}{\delta \rho_s \left( X \right)}
    G^{\text{($ m $)}}_{\lambda, \, s_1 \ldots s_m} \left( X_1, \ldots, X_m \right)
    =
    G^{\text{($ m + 1 $)}}_{\lambda, \, s_1 \ldots s_{m + 1}} \left( X_1, \ldots, X_{m + 1} \right).
  \end{equation}
  This is obtained by differentiating Eqs.~\eqref{eq:gamma1} and \eqref{eq:gdef}:
  \begin{align}
    & \frac{\delta}{\delta \rho_s \left( X \right)}
      G^{\text{($ m $)}}_{\lambda, \, s_1 \ldots s_m} \left( X_1, \ldots, X_m \right)
      =
      \frac{\delta}{\delta \rho_s \left( X \right)}
      \frac{\delta^m W_{\lambda}
      \left[
      J_{\text{sup}, \, \lambda, \, {\uparrow}} \left[ \rhou, \rhod \right],
      J_{\text{sup}, \, \lambda, \, {\downarrow}} \left[ \rhou, \rhod \right]
      \right]}
      {\delta J_{s_1} \left( X_1 \right) \, \cdots \, \delta J_{s_m} \left( X_m \right)}
      \notag \\
    & =
      \sum_{s_{m + 1}}
      \int_{X_{m + 1}}
      \frac{\delta J_{\text{sup}, \, \lambda, \, s_{m + 1}} \left[ \rhou, \rhod \right] \left( X_{m + 1} \right)}
      {\delta \rho_s \left( X \right)}
      \frac{\delta^{m + 1} W_{\lambda}
      \left[
      J_{\text{sup}, \, \lambda, \, {\uparrow}} \left[ \rhou, \rhod \right],
      J_{\text{sup}, \, \lambda, \, {\downarrow}} \left[ \rhou, \rhod \right]
      \right]}
      {\delta J_{s_1} \left( X_1 \right) \, \cdots \, \delta J_{s_m} \left( X_m \right) \, \delta J_{s_{m + 1}} \left( X_{m + 1} \right)}
      \notag \\
    & =
      \sum_{s_{m + 1}}
      \int_{X_{m + 1}}
      \frac{\delta^2 \Gamma_{\lambda} \left[ \rhou, \rhod \right]}{\delta \rho_s \left( X \right) \, \delta \rho_{s_{m + 1}} \left( X_{m + 1} \right)}
      G^{\text{($ m + 1 $)}}_{\lambda, \, s_1 \ldots s_{m + 1}} \left( X_1, \ldots, X_{m + 1} \right).
  \end{align}
  Multiplying
  $ \Gamma^{\text{($ 2 $)} {-1}}_{\lambda} $
  to both sides and using Eq.~\eqref{eq:gam2g2}, we have Eq.~\eqref{eq:gm1}.
  \par
  The derivation of Eq.~\eqref{eq:flown} is based on the mathematical induction.
  As a first step, we derive the equation for $ m = 1 $.
  Differentiating Eq.~\eqref{eq:flow_eq}, we have
  \begin{align}
    \partial_{\lambda}
    \frac{\delta \Gamma_{\lambda} \left[ \rhou, \rhod \right]}{\delta \rho_{s_1} \left( X_1 \right)}
    = & \,
        \frac{1}{2}
        \iint_{X, \, X'}
        U_{\text{2b}} \left( X - X' \right)
        \left[
        \rho_{\Delta}\left( X \right) \delta \left( X' - X_1 \right)
        +
        \rho_{\Delta}\left( X' \right) \delta \left( X - X_1 \right)
        \right]
        \notag \\
      & \,
        +
        \frac{1}{2}
        \iint_{X, \, X'}
        U_{\text{2b}} \left( X - X' \right)
        \left[
        \sum_{s, \, s'}
        \frac{
        \delta G^{\text{($ 2 $)}}_{\lambda, \, s s'} \left( X_{\epsilon'}, X' \right)}
        {\delta \rho_{s_1} \left( X_1 \right)}
        -
        \delta \left( X - X_1 \right) \delta \left( \bm{x} - \bm{x}' \right)
        \right].
  \end{align}
  Multiplying
  $ G^{\text{($ 2 $)}}_{\lambda} $ and using
  Eqs.~\eqref{eq:gm1} and \eqref{eq:gamma1}, we obtain
  \begin{align}
    & \sum_s
      \int_X
      G^{\text{($ 2 $)}}_{\lambda, \, s_1 s} \left( X_1, X \right)
      \partial_{\lambda}
      J_{\text{sup}, \, \lambda, \, s} \left[ \rhou, \rhod \right] \left( X \right)
      \notag \\
    = & \,
        \frac{1}{2}
        \sum_s
        \iint_{X, \, X'}
        U_{\text{2b}} \left( X - X' \right)
        \left[
        \rho_{\Delta} \left( X \right)
        G^{\text{($ 2 $)}}_{\lambda, \, s s_1} \left( X', X_1 \right)
        +
        \rho_{\Delta} \left( X' \right)
        G^{\text{($ 2 $)}}_{\lambda, \, s s_1} \left( X, X_1 \right)
        \right]
        \notag \\
    & \,
      +
      \frac{1}{2}
      \iint_{X, \, X'}
      U_{\text{2b}} \left( X - X' \right)
      \left[
      \sum_{s \, s'}
      G^{\text{($ 3 $)}}_{\lambda, \, s s' s_1} \left( X_{\epsilon'}, X', X_1 \right)
      -
      \sum_s
      \delta \left( \bm{x} - \bm{x}' \right)
      G^{\text{($ 2 $)}}_{\lambda, \, s s_1} \left( X, X_1 \right)
      \right].
      \label{eq:m1res}
  \end{align}
  Remembering
  $ \partial_{\lambda} G^{\text{($ 1 $)}}_{\lambda, \, s} \left( X_1 \right) = \partial_{\lambda} \rho_s \left( X_1 \right) = 0 $,
  one finds that Eq.~\eqref{eq:m1res} is equivalent to
  Eq.~\eqref{eq:flown} in the case of $ m = 1 $.
  \par
  Next, we assume that Eq.~\eqref{eq:flown} holds for $ m = j $
  and derive the equation for $ m = j + 1 $.
  Differentiating Eq.~\eqref{eq:flown} for $ m = j $
  with respect to the density, multiplying
  $ G^{\text{($ 2 $)}}_{\lambda} $,
  and using Eq.~\eqref{eq:gm1}, we obtain
  \begin{align}
    & \sum_s
      \int_X
      G^{\text{($ 2 $)}}_{\lambda s_{j + 1} s} \left( X_{j + 1}, X \right)
      \partial_{\lambda}
      \frac{\delta}{\delta \rho_s \left( X \right)}
      G^{\text{($ j $)}}_{\lambda, \, s_1 \ldots s_j} \left( X_1, \ldots X_j \right)
      \notag \\
    & =
      \sum_{s \, s'}
      \iint_{X, \, X'}
      G^{\text{($ j + 1 $)}}_{\lambda, \, s s_1 \ldots s_j} \left( X, X_1, \ldots X_j \right)
      G^{\text{($ 2 $)}}_{\lambda, \,  s_{j + 1} s'} \left( X_{j + 1}, X' \right)
      \partial_{\lambda}
      \frac{\delta J_{\text{sup}, \, \lambda, \, s} \left[ \rhou, \rhod \right] \left( X \right)}
      {\delta \rho_{s'} \left( X' \right)}
      \notag \\
    & \quad
      +
      \sum_s
      \int_X
      G^{\text{($ j + 2 $)}}_{\lambda, \, s s_1 \ldots s_{j + 1}} \left( X, X_1, \ldots X_{j + 1} \right)
      \partial_{\lambda} J_{\text{sup}, \, \lambda, \, s} \left[ \rhou, \rhod \right] \left( X \right)
      \notag \\
    & \quad
      -
      \frac{1}{2}
      \sum_{s \, s'}
      \iint_{X, \, X'}
      U_{\text{2b}} \left( X, X' \right)
      \notag \\
    & \qquad
      \times
      \left(
      G^{\text{($ 2 $)}}_{\lambda, \, s' s_{j + 1}} \left( X, X_{j + 1} \right)
      G^{\text{($ j + 1 $)}}_{\lambda, \, s s_1 \ldots s_j} \left( X', X_1, \ldots X_j \right)
      +
      G^{\text{($ 2 $)}}_{\lambda, \, s' s_{j + 1}} \left( X', X_{j + 1} \right)
      G^{\text{($ j + 1 $)}}_{\lambda, \, s s_1 \ldots s_j} \left( X, X_1, \ldots X_j \right)
      \right)
      \notag \\
    & \quad
      -
      \frac{1}{2}
      \sum_s
      \iint_{X, \, X'}
      U_{\text{2b}} \left( X, X' \right)
      \left(
      \rho_{\Delta} \left( X \right)
      G^{\text{($ j + 2 $)}}_{\lambda, \, s s_1 \ldots s_{j + 1}} \left( X', X_1, \ldots X_{j + 1} \right)
      +
      \rho_{\Delta} \left( X' \right)
      G^{\text{($ j + 2 $)}}_{\lambda, \, s s_1 \ldots s_{j + 1}} \left( X, X_1, \ldots X_{j + 1} \right)
      \right)
      \notag \\
    & \quad
      -
      \frac{1}{2}
      \sum_{s \, s'}
      \iint_{X, \, X'}
      U_{\text{2b}} \left( X, X' \right)
      \left(
      G^{\text{($ j + 2 $)}}_{\lambda, \, s s' s_1 \ldots s_{j + 1}} \left( X_{\epsilon'}, X', X_1, \ldots X_{j + 1} \right)
      -
      G^{\text{($ j + 2 $)}}_{\lambda, \, s s_1 \ldots s_{j + 1}} \left( X, X_1, \ldots X_{j + 1} \right)
      \delta_{s s'}
      \delta \left( \bm{x} -\bm{x}' \right)
      \right.
      \notag \\
    & \qquad
      \left.
      +
      \sum_{k = 1}^{j - 1}
      \frac{1}{k! \, \left( j - k \right)!}
      \sum_{\sigma \in S_j}
      \left.
      G^{\text{($ k + 2 $)}}_{\lambda, \, s s_{\sigma \left( 1 \right)} \ldots s_{\sigma \left( k \right)} s_{j + 1}}
      \left( X, X_{\sigma \left( 1 \right)}, \ldots, X_{\sigma \left( k \right)}, X_{j + 1} \right)
      G^{\text{($ j - k + 1 $)}}_{\lambda, \, s' s_{\sigma \left( k + 1 \right)} \ldots s_{\sigma \left( j \right)}}
      \left( X', X_{\sigma \left( k + 1 \right)}, \ldots, X_{\sigma \left( j \right)} \right)
      \right.
      \right.
      \notag \\
    & \qquad
      +
      \left.
      \left.
      G^{\text{($ k + 1 $)}}_{\lambda, \, s s_{\sigma \left( 1 \right)} \ldots s_{\sigma \left( k \right)}}
      \left( X, X_{\sigma \left( 1 \right)}, \ldots, X_{\sigma \left( k \right)} \right)
      G^{\text{($ j - k + 2 $)}}_{\lambda, \, s' s_{\sigma \left( k + 1 \right)} \ldots s_{\sigma \left( j \right) s_{j + 1}}}
      \left( X', X_{\sigma \left( k + 1 \right)}, \ldots, X_{\sigma \left( j \right)}, X_{j + 1} \right)
      \right]
      \right).
      \label{eq:flowj1}
  \end{align}
  The left-hand side is evaluated through the derivative of Eq.~\eqref{eq:gm1}
  with respect to $ \lambda $:
  \begin{align}
    & \sum_s
      \int_X
      G^{\text{($ 2 $)}}_{\lambda, \, s_{j + 1} s} \left( X_{j + 1}, X \right)
      \partial_{\lambda}
      \frac{\delta}{\delta \rho_s \left( X \right)}
      G^{\text{($ j $)}}_{\lambda, \, s_1 \ldots s_j} \left( X_1, \ldots, X_j \right)
      \notag \\
    & =
      \partial_{\lambda}
      G^{\text{($ j + 1 $)}}_{\lambda, \, s_1 \ldots s_{j + 1}} \left( X_1, \ldots, X_{j + 1} \right)
      \sum_s
      \int_X
      \partial_{\lambda}
      G^{\text{($ 2 $)}}_{\lambda, \, s_{j + 1} s} \left( X_{j + 1}, X \right)
      \frac{\delta}{\delta \rho_s \left( X \right)}
      G^{\text{($ j $)}}_{\lambda, \, s_1 \ldots s_j} \left( X_1, \ldots, X_j \right).
  \end{align}
  By use of Eq.~\eqref{eq:jrw2}, this is rewritten as follows:
  \begin{align}
    & \sum_s
      \int_X
      G^{\text{($ 2 $)}}_{\lambda, \, s_{j + 1} s} \left( X_{j + 1}, X \right)
      \partial_{\lambda}
      \frac{\delta}{\delta \rho_s \left( X \right)}
      G^{\text{($ j $)}}_{\lambda, \, s_1 \ldots s_j} \left( X_1, \ldots, X_j \right)
      \notag \\
    & =
      \partial_{\lambda}
      G^{\text{($ j + 1 $)}}_{\lambda, \, s_1 \ldots s_{j + 1}} \left( X_1, \ldots, X_{j + 1} \right)
      \notag \\
    & \quad
      -
      \sum_{s \, s'}
      \iint_{X, \, X'}
      \partial_{\lambda}
      G^{\text{($ 2 $)}}_{\lambda, \, s_{j + 1} s} \left( X_{j + 1}, X \right)
      \sum_t
      \int_Y
      \frac{\delta J_{\text{sup}, \, \lambda, \, t} \left[ \rhou, \rhod \right] \left( Y \right)}
      {\delta \rho_s \left( X \right)}
      G^{\text{($ 2 $)}}_{\lambda, \, t s'} \left( Y, X' \right)
      \frac{\delta}{\delta \rho_{s'} \left( X' \right)}
      G^{\text{($ j $)}}_{\lambda, \, s_1 \ldots s_j} \left( X_1, \ldots, X_j \right).
      \label{eq:flowjlhs1}
  \end{align}
  By differentiating Eq.~\eqref{eq:jrw2} with respect to $ \lambda $, we have
  \begin{equation}
    \sum_s
    \int_X
    \partial_{\lambda}
    G^{\text{($ 2 $)}}_{\lambda, \, s_{j + 1} s} \left( X_{j + 1}, X \right)
    \frac{\delta J_{\text{sup}, \, \lambda, \, t} \left[ \rhou, \rhod \right] \left( Y \right)}{\delta \rho_s \left( X \right)}
    =
    -
    \sum_s
    \int_X
    G^{\text{($ 2 $)}}_{\lambda, \, s_{j + 1} s} \left( X_{j + 1}, X \right)
    \partial_{\lambda}
    \frac{\delta J_{\text{sup}, \, \lambda, \, t} \left[ \rhou, \rhod \right] \left( Y \right)}{\delta \rho_s \left( X \right)}.
  \end{equation}
  By use of this relation and Eq.~\eqref{eq:gm1},
  Eq.~\eqref{eq:flowjlhs1} is rewritten as follows:
  \begin{align}
    & \sum_s
      \int_X
      G^{\text{($ 2 $)}}_{\lambda, \, s_{j + 1} s} \left( X_{j + 1}, X \right)
      \partial_{\lambda}
      \frac{\delta}{\delta \rho_s \left( X \right)}
      G^{\text{($ j $)}}_{\lambda, \, s_1 \ldots s_j} \left( X_1, \ldots, X_j \right)
      \notag \\
    & =
      \partial_{\lambda}
      G^{\text{($ j + 1 $)}}_{\lambda, \, s_1 \ldots s_{j + 1}} \left( X_1, \ldots, X_{j + 1} \right)
      +
      \sum_{s \, t}
      \iint_{X, \, Y}
      G^{\text{($ j + 1 $)}}_{\lambda, \, t s_1 \ldots s_j} \left( Y, X_1, \ldots, X_j \right)
      G^{\text{($ 2 $)}}_{\lambda, \, s_{j + 1} s} \left( X_{j + 1}, X \right)
      \partial_{\lambda}
      \frac{\delta J_{\text{sup}, \, \lambda, \, t} \left[ \rhou, \rhod \right] \left( Y \right)}{\delta \rho_s \left( X \right)}.
      \label{eq:flowjlhsfinal}
  \end{align}
  The second term of this equation and
  the first term in the right-hand side of Eq.~\eqref{eq:flowj1}
  cancel each other.
  The last term in the right-hand side of Eq.~\eqref{eq:flowj1} is deformed as
  follows:
  \begin{align}
    & \sum_{k = 1}^{j - 1}
      \frac{1}{k! \, \left(j-k\right)!}
      \sum_{\sigma \in S_{j}}
      \left[
      G^{\text{($ k + 2 $)}}_{\lambda, \, s s_{\sigma \left( 1 \right)} \ldots \ldots s_{\sigma \left( k \right)} s_{j + 1}}
      \left( X, X_{\sigma \left( 1 \right)}, \ldots, X_{\sigma \left( k \right)}, X_{j + 1} \right)
      G^{\text{($ j - k + 1 $)}}_{\lambda, \, s' s_{\sigma \left( k + 1 \right)} \ldots \ldots s_{\sigma \left( j \right)}}
      \left( X', X_{\sigma \left( k + 1 \right)}, \ldots, X_{\sigma \left( j \right)} \right)
      \right.
      \notag \\
    & \quad
      +
      \left.
      G^{\text{($ k + 1 $)}}_{\lambda, \, s s_{\sigma \left( 1 \right)} \ldots \ldots s_{\sigma \left( k \right)}}
      \left( X, X_{\sigma \left( 1 \right)}, \ldots, X_{\sigma \left( k \right)} \right)
      G^{\text{($ j - k + 2 $)}}_{\lambda, \, s' s_{\sigma \left( k + 1 \right)} \ldots \ldots s_{\sigma \left( j \right)} s_{j + 1}}
      \left( X', X_{\sigma \left( k + 1 \right)}, \ldots, X_{\sigma \left( j \right)}, X_{j + 1} \right)
      \right]
      \notag \\
    & =
      \sum_{\sigma \in S_{j}}
      \left[
      \sum_{k = 2}^j
      \frac{1}{\left( k -1 \right)! \, \left( j - k + 1 \right)!}
      G^{\text{($ k + 1 $)}}_{\lambda, \, s s_{\sigma \left( 1 \right)} \ldots \ldots s_{\sigma \left( k - 1 \right)} s_{j + 1}}
      \left( X, X_{\sigma \left( 1 \right)}, \ldots, X_{\sigma \left( k - 1 \right)}, X_{j + 1}\right)
      G^{\text{($ j - k + 2 $)}}_{\lambda, \, s' s_{\sigma \left( k \right)} \ldots \ldots s_{\sigma \left( j \right)}}
      \left( X', X_{\sigma \left( k \right)}, \ldots, X_{\sigma \left( j \right)} \right)
      \right.
      \notag \\
    & \quad
      \left.
      +
      \sum_{k = 1}^{j - 1}
      \frac{1}{k! \, \left( j - k \right)!}
      G^{\text{($ k + 1 $)}}_{\lambda, \, s s_{\sigma \left( 1 \right)} \ldots \ldots s_{\sigma \left( k \right)}}
      \left( X, X_{\sigma \left( 1 \right)}, \ldots, X_{\sigma \left( k \right)} \right)
      G^{\text{($ j - k + 2 $)}}_{\lambda, \, s' s_{\sigma \left( k + 1 \right)} \ldots \ldots s_{\sigma \left( j \right)} s_{j + 1}}
      \left( X', X_{\sigma \left( k + 1 \right)}, \ldots, X_{\sigma \left( j \right)}, X_{j + 1} \right)
      \right]
      \notag \\
    & =
      \sum_{\sigma \in S_j}
      \frac{1}{k! \, \left(j + 1 -k \right)!}
      \left[
      \sum_{k = 2}^j
      k
      G^{\text{($ k + 1 $)}}_{\lambda, \, s s_{\sigma \left( 1 \right)} \ldots \ldots s_{\sigma \left( k - 1 \right)} s_{j + 1}}
      \left( X, X_{\sigma \left( 1 \right)}, \ldots, X_{\sigma \left( k - 1 \right)}, X_{j + 1} \right)
      G^{\text{($ j - k + 2 $)}}_{\lambda, \, s' s_{\sigma \left( k \right)} \ldots \ldots s_{\sigma \left( j \right)}}
      \left( X', X_{\sigma \left( k \right)}, \ldots, X_{\sigma \left( j \right)} \right)
      \right.
      \notag \\
    & \quad
      \left.
      +
      \sum_{k = 1}^{j - 1}
      \left(j + 1 - k \right)
      G^{\text{($ k + 1 $)}}_{\lambda, \, s s_{\sigma \left( 1 \right)} \ldots \ldots s_{\sigma \left( k \right)}}
      \left( X, X_{\sigma \left( 1 \right)}, \ldots, X_{\sigma \left( k \right)} \right)
      G^{\text{($ j - k + 2 $)}}_{\lambda, \, s' s_{\sigma \left( k + 1 \right)} \ldots \ldots s_{\sigma \left( j \right)} s_{j + 1}}
      \left( X', X_{\sigma \left( k + 1 \right)}, \ldots, X_{\sigma \left( j \right)}, X_{j + 1} \right)
      \right]
      \notag \\
    & =
      \sum_{\sigma \in S_j}
      \frac{1}{k! \, \left( j + 1 - k \right)!}	
      \left[
      \sum_{k = 1}^j
      k
      G^{\text{($ k + 1 $)}}_{\lambda, \, s s_{\sigma \left( 1 \right)} \ldots \ldots s_{\sigma \left( k - 1 \right)} s_{j + 1}}
      \left( X, X_{\sigma \left( 1 \right)}, \ldots, X_{\sigma \left( k - 1 \right)}, X_{j + 1} \right)
      G^{\text{($ j - k + 2 $)}}_{\lambda, \, s' s_{\sigma \left( k \right)} \ldots \ldots s_{\sigma \left( j \right)}}
      \left( X', X_{\sigma \left( k \right)}, \ldots, X_{\sigma \left( j \right)} \right)
      \right.
      \notag \\
    & \qquad
      \left.
      +
      \sum_{k = 1}^{j - 1}
      \left(j + 1 - k \right)
      G^{\text{($ k + 1 $)}}_{\lambda, \, s s_{\sigma \left( 1 \right)} \ldots \ldots s_{\sigma \left( k \right)}}
      \left( X, X_{\sigma \left( 1 \right)}, \ldots, X_{\sigma \left( k \right)} \right)
      G^{\text{($ j - k + 2 $)}}_{\lambda, \, s' s_{\sigma \left( k + 1 \right)} \ldots \ldots s_{\sigma \left( j \right)} s_{j + 1}}
      \left( X', X_{\sigma \left( k + 1 \right)}, \ldots, X_{\sigma \left( j \right)}, X_{j + 1} \right)
      \right]
      \notag \\
    & \quad
      -
      \sum_{\sigma \in S_j}
      \frac{1}{j!}
      \left[
      G^{\text{($ 2 $)}}_{\lambda, \, s s_{j + 1}} \left( X, X_{j + 1} \right)
      G^{\text{($ j + 1 $)}}_{\lambda, \, s' s_{\sigma \left( 1 \right)} \ldots s_{\sigma \left( j \right)}}
      \left( X', X_{\sigma \left( 1 \right)}, \ldots, X_{\sigma \left( j \right)} \right)
      \right.
      \notag \\
    & \qquad
      \left.
      +
      G^{\text{($ j + 1 $)}}_{\lambda, \, s s_{\sigma \left( 1 \right)} \ldots s_{\sigma \left( j \right)}}
      \left( X, X_{\sigma \left( 1 \right)}, \ldots, X_{\sigma \left( j \right)} \right)
      G^{\text{($ 2 $)}}_{\lambda, \, s' s_{j + 1}} \left( X', X_{j + 1} \right)
      \right]
      \notag \\
    & =
      \sum_{\sigma \in S_{j+1}}
      \sum_{k = 1}^j
      \frac{1}{k! \, \left( j + 1 - k \right)!}
      G^{\text{($ k + 1 $)}}_{\lambda, \, s s_{\sigma \left( 1 \right)} \ldots s_{\sigma \left( k \right)}}
      \left( X, X_{\sigma \left( 1 \right)}, \ldots, X_{\sigma \left( k \right)} \right)
      G^{\text{($ j - k + 2 $)}}_{\lambda, \, s' s_{\sigma \left( k + 1 \right)} \ldots s_{\sigma \left( j + 1 \right)}}
      \left( X', X_{\sigma \left( k + 1 \right)}, \ldots, X_{\sigma \left( j + 1 \right)} \right)
      \notag \\
    & \quad
      -
      G^{\text{($ 2 $)}}_{\lambda, \, s s_{j + 1}} \left( X, X_{j + 1} \right)
      G^{\text{($ j + 1 $)}}_{\lambda, \, s' s_1 \ldots  s_j} \left( X', X_1, \ldots, X_j \right)
      -
      G^{\text{($ j + 1 $)}}_{\lambda, \, s s_1 \ldots  s_j} \left( X, X_1, \ldots, X_j \right)
      G^{\text{($ 2 $)}}_{\lambda, \, s' s_{j + 1}} \left( X', X_{j + 1} \right).
      \label{eq:flowjrhslast}
  \end{align}
  By substituting Eqs.~\eqref{eq:flowjlhsfinal} and \eqref{eq:flowjrhslast}
  into Eq.~\eqref{eq:flowj1},
  we obtain Eq.~\eqref{eq:flown} for $ m = j + 1 $.
  Therefore, Eq.~\eqref{eq:flown} holds for all integers $ m \geq 1 $.
\end{widetext}
% 
% \bibliography{./frgdft_3DHEG_spin_ref}

\begin{thebibliography}{65}%
\makeatletter
\providecommand \@ifxundefined [1]{%
 \@ifx{#1\undefined}
}%
\providecommand \@ifnum [1]{%
 \ifnum #1\expandafter \@firstoftwo
 \else \expandafter \@secondoftwo
 \fi
}%
\providecommand \@ifx [1]{%
 \ifx #1\expandafter \@firstoftwo
 \else \expandafter \@secondoftwo
 \fi
}%
\providecommand \natexlab [1]{#1}%
\providecommand \enquote  [1]{``#1''}%
\providecommand \bibnamefont  [1]{#1}%
\providecommand \bibfnamefont [1]{#1}%
\providecommand \citenamefont [1]{#1}%
\providecommand \href@noop [0]{\@secondoftwo}%
\providecommand \href [0]{\begingroup \@sanitize@url \@href}%
\providecommand \@href[1]{\@@startlink{#1}\@@href}%
\providecommand \@@href[1]{\endgroup#1\@@endlink}%
\providecommand \@sanitize@url [0]{\catcode `\\12\catcode `\$12\catcode
  `\&12\catcode `\#12\catcode `\^12\catcode `\_12\catcode `\%12\relax}%
\providecommand \@@startlink[1]{}%
\providecommand \@@endlink[0]{}%
\providecommand \url  [0]{\begingroup\@sanitize@url \@url }%
\providecommand \@url [1]{\endgroup\@href {#1}{\urlprefix }}%
\providecommand \urlprefix  [0]{URL }%
\providecommand \Eprint [0]{\href }%
\providecommand \doibase [0]{https://doi.org/}%
\providecommand \selectlanguage [0]{\@gobble}%
\providecommand \bibinfo  [0]{\@secondoftwo}%
\providecommand \bibfield  [0]{\@secondoftwo}%
\providecommand \translation [1]{[#1]}%
\providecommand \BibitemOpen [0]{}%
\providecommand \bibitemStop [0]{}%
\providecommand \bibitemNoStop [0]{.\EOS\space}%
\providecommand \EOS [0]{\spacefactor3000\relax}%
\providecommand \BibitemShut  [1]{\csname bibitem#1\endcsname}%
\let\auto@bib@innerbib\@empty
%</preamble>
\bibitem [{\citenamefont {Hohenberg}\ and\ \citenamefont {Kohn}(1964)}]{hoh64}%
  \BibitemOpen
  \bibfield  {author} {\bibinfo {author} {\bibfnamefont {P.}~\bibnamefont
  {Hohenberg}}\ and\ \bibinfo {author} {\bibfnamefont {W.}~\bibnamefont
  {Kohn}},\ }\bibfield  {title} {\bibinfo {title} {{Inhomogeneous Electron
  Gas}},\ }\href {https://link.aps.org/doi/10.1103/PhysRev.136.B864} {\bibfield
   {journal} {\bibinfo  {journal} {Phys. Rev.}\ }\textbf {\bibinfo {volume}
  {136}},\ \bibinfo {pages} {B864} (\bibinfo {year} {1964})}\BibitemShut
  {NoStop}%
\bibitem [{\citenamefont {Kohn}\ and\ \citenamefont {Sham}(1965)}]{koh65}%
  \BibitemOpen
  \bibfield  {author} {\bibinfo {author} {\bibfnamefont {W.}~\bibnamefont
  {Kohn}}\ and\ \bibinfo {author} {\bibfnamefont {L.~J.}\ \bibnamefont
  {Sham}},\ }\bibfield  {title} {\bibinfo {title} {{Self-Consistent Equations
  Including Exchange and Correlation Effects}},\ }\href
  {https://link.aps.org/doi/10.1103/PhysRev.140.A1133} {\bibfield  {journal}
  {\bibinfo  {journal} {Phys. Rev.}\ }\textbf {\bibinfo {volume} {140}},\
  \bibinfo {pages} {A1133} (\bibinfo {year} {1965})}\BibitemShut {NoStop}%
\bibitem [{\citenamefont {Kohn}(1999)}]{koh99}%
  \BibitemOpen
  \bibfield  {author} {\bibinfo {author} {\bibfnamefont {W.}~\bibnamefont
  {Kohn}},\ }\bibfield  {title} {\bibinfo {title} {{Nobel Lecture: Electronic
  structure of matter---wave functions and density functionals}},\ }\href
  {https://link.aps.org/doi/10.1103/RevModPhys.71.1253} {\bibfield  {journal}
  {\bibinfo  {journal} {Rev. Mod. Phys.}\ }\textbf {\bibinfo {volume} {71}},\
  \bibinfo {pages} {1253} (\bibinfo {year} {1999})}\BibitemShut {NoStop}%
\bibitem [{\citenamefont {Perdew}\ and\ \citenamefont {Schmidt}(2001)}]{per01}%
  \BibitemOpen
  \bibfield  {author} {\bibinfo {author} {\bibfnamefont {J.~P.}\ \bibnamefont
  {Perdew}}\ and\ \bibinfo {author} {\bibfnamefont {K.}~\bibnamefont
  {Schmidt}},\ }\bibfield  {title} {\bibinfo {title} {{Jacob's ladder of
  density functional approximations for the exchange-correlation energy}},\
  }\href {https://aip.scitation.org/doi/abs/10.1063/1.1390175} {\bibfield
  {journal} {\bibinfo  {journal} {AIP Conf. Proc.}\ }\textbf {\bibinfo {volume}
  {577}},\ \bibinfo {pages} {1} (\bibinfo {year} {2001})}\BibitemShut {NoStop}%
\bibitem [{\citenamefont {Fukuda}\ \emph {et~al.}(1994)\citenamefont {Fukuda},
  \citenamefont {Kotani}, \citenamefont {Suzuki},\ and\ \citenamefont
  {Yokojima}}]{fuk94}%
  \BibitemOpen
  \bibfield  {author} {\bibinfo {author} {\bibfnamefont {R.}~\bibnamefont
  {Fukuda}}, \bibinfo {author} {\bibfnamefont {T.}~\bibnamefont {Kotani}},
  \bibinfo {author} {\bibfnamefont {Y.}~\bibnamefont {Suzuki}},\ and\ \bibinfo
  {author} {\bibfnamefont {S.}~\bibnamefont {Yokojima}},\ }\bibfield  {title}
  {\bibinfo {title} {{Density Functional Theory through Legendre
  Transformation}},\ }\href {https://doi.org/10.1143/ptp/92.4.833} {\bibfield
  {journal} {\bibinfo  {journal} {Prog. Theor. Phys.}\ }\textbf {\bibinfo
  {volume} {92}},\ \bibinfo {pages} {833} (\bibinfo {year} {1994})}\BibitemShut
  {NoStop}%
\bibitem [{\citenamefont {Fukuda}\ \emph {et~al.}(1995)\citenamefont {Fukuda},
  \citenamefont {Komachiya}, \citenamefont {Yokojima}, \citenamefont {Suzuki},
  \citenamefont {Okumura},\ and\ \citenamefont {Inagaki}}]{fuk95}%
  \BibitemOpen
  \bibfield  {author} {\bibinfo {author} {\bibfnamefont {R.}~\bibnamefont
  {Fukuda}}, \bibinfo {author} {\bibfnamefont {M.}~\bibnamefont {Komachiya}},
  \bibinfo {author} {\bibfnamefont {S.}~\bibnamefont {Yokojima}}, \bibinfo
  {author} {\bibfnamefont {Y.}~\bibnamefont {Suzuki}}, \bibinfo {author}
  {\bibfnamefont {K.}~\bibnamefont {Okumura}},\ and\ \bibinfo {author}
  {\bibfnamefont {T.}~\bibnamefont {Inagaki}},\ }\bibfield  {title} {\bibinfo
  {title} {{Novel use of Legendre transformation in field theory and many
  particle systems: On-shell expansion and inversion method}},\ }\href
  {https://doi.org/10.1143/PTPS.121.1} {\bibfield  {journal} {\bibinfo
  {journal} {Prog. Theor. Phys. Suppl.}\ }\textbf {\bibinfo {volume} {121}},\
  \bibinfo {pages} {1} (\bibinfo {year} {1995})}\BibitemShut {NoStop}%
\bibitem [{\citenamefont {Valiev}\ and\ \citenamefont
  {Fernando}(1997)}]{val97}%
  \BibitemOpen
  \bibfield  {author} {\bibinfo {author} {\bibfnamefont {M.}~\bibnamefont
  {Valiev}}\ and\ \bibinfo {author} {\bibfnamefont {G.~W.}\ \bibnamefont
  {Fernando}},\ }\bibfield  {title} {\bibinfo {title} {{Generalized Kohn-Sham
  Density-Functional Theory via Effective Action Formalism}},\ }\Eprint
  {https://arxiv.org/abs/cond-mat/9702247} {arXiv:cond-mat/9702247}  (\bibinfo
  {year} {1997})\BibitemShut {NoStop}%
\bibitem [{\citenamefont {Furnstahl}(2020)}]{Furnstahl2020Eur.Phys.J.A56_85}%
  \BibitemOpen
  \bibfield  {author} {\bibinfo {author} {\bibfnamefont {R.~J.}\ \bibnamefont
  {Furnstahl}},\ }\bibfield  {title} {\bibinfo {title} {{Turning the nuclear
  energy density functional method into a proper effective field theory:
  reflections}},\ }\href {https://doi.org/10.1140/epja/s10050-020-00095-y}
  {\bibfield  {journal} {\bibinfo  {journal} {Eur. Phys. J. A}\ }\textbf
  {\bibinfo {volume} {56}},\ \bibinfo {pages} {85} (\bibinfo {year}
  {2020})}\BibitemShut {NoStop}%
\bibitem [{\citenamefont {Polonyi}\ and\ \citenamefont {Sailer}(2002)}]{pol02}%
  \BibitemOpen
  \bibfield  {author} {\bibinfo {author} {\bibfnamefont {J.}~\bibnamefont
  {Polonyi}}\ and\ \bibinfo {author} {\bibfnamefont {K.}~\bibnamefont
  {Sailer}},\ }\bibfield  {title} {\bibinfo {title} {{Effective actions and the
  density functional theory}},\ }\href
  {https://doi.org/10.1103/PhysRevB.66.155113} {\bibfield  {journal} {\bibinfo
  {journal} {Phys. Rev. B}\ }\textbf {\bibinfo {volume} {66}},\ \bibinfo
  {pages} {155113} (\bibinfo {year} {2002})}\BibitemShut {NoStop}%
\bibitem [{\citenamefont {Schwenk}\ and\ \citenamefont
  {Polonyi}(2004)}]{sch04}%
  \BibitemOpen
  \bibfield  {author} {\bibinfo {author} {\bibfnamefont {A.}~\bibnamefont
  {Schwenk}}\ and\ \bibinfo {author} {\bibfnamefont {J.}~\bibnamefont
  {Polonyi}},\ }\bibfield  {title} {\bibinfo {title} {{Towards density
  functional calculations from nuclear forces}},\ }in\ \href
  {http://theory.gsi.de/hirschegg/2004/Proceedings/Schwenk.ps} {\emph {\bibinfo
  {booktitle} {{32nd International Workshop on Gross Properties of Nuclei and
  Nuclear Excitation: Probing Nuclei and Nucleons with Electrons and Photons
  (Hirschegg 2004) Hirschegg, Austria, January 11-17, 2004}}}}\ (\bibinfo
  {year} {2004})\ pp.\ \bibinfo {pages} {273--282},\ \Eprint
  {https://arxiv.org/abs/nucl-th/0403011} {arXiv:nucl-th/0403011} \BibitemShut
  {NoStop}%
\bibitem [{\citenamefont {Wegner}\ and\ \citenamefont
  {Houghton}(1973)}]{weg73}%
  \BibitemOpen
  \bibfield  {author} {\bibinfo {author} {\bibfnamefont {F.~J.}\ \bibnamefont
  {Wegner}}\ and\ \bibinfo {author} {\bibfnamefont {A.}~\bibnamefont
  {Houghton}},\ }\bibfield  {title} {\bibinfo {title} {{Renormalization Group
  Equation for Critical Phenomena}},\ }\href
  {https://doi.org/10.1103/PhysRevA.8.401} {\bibfield  {journal} {\bibinfo
  {journal} {Phys. Rev. A}\ }\textbf {\bibinfo {volume} {8}},\ \bibinfo {pages}
  {401} (\bibinfo {year} {1973})}\BibitemShut {NoStop}%
\bibitem [{\citenamefont {Wilson}\ and\ \citenamefont {Kogut}(1974)}]{wil74}%
  \BibitemOpen
  \bibfield  {author} {\bibinfo {author} {\bibfnamefont {K.~G.}\ \bibnamefont
  {Wilson}}\ and\ \bibinfo {author} {\bibfnamefont {J.}~\bibnamefont {Kogut}},\
  }\bibfield  {title} {\bibinfo {title} {{The renormalization group and the $
  \epsilon $ expansion}},\ }\href
  {https://doi.org/https://doi.org/10.1016/0370-1573(74)90023-4} {\bibfield
  {journal} {\bibinfo  {journal} {Phys. Rep.}\ }\textbf {\bibinfo {volume}
  {12}},\ \bibinfo {pages} {75 } (\bibinfo {year} {1974})}\BibitemShut
  {NoStop}%
\bibitem [{\citenamefont {Polchinski}(1984)}]{pol84}%
  \BibitemOpen
  \bibfield  {author} {\bibinfo {author} {\bibfnamefont {J.}~\bibnamefont
  {Polchinski}},\ }\bibfield  {title} {\bibinfo {title} {{Renormalization and
  effective lagrangians}},\ }\href
  {https://doi.org/https://doi.org/10.1016/0550-3213(84)90287-6} {\bibfield
  {journal} {\bibinfo  {journal} {Nucl. Phys. B}\ }\textbf {\bibinfo {volume}
  {231}},\ \bibinfo {pages} {269 } (\bibinfo {year} {1984})}\BibitemShut
  {NoStop}%
\bibitem [{\citenamefont {Wetterich}(1993)}]{wet93}%
  \BibitemOpen
  \bibfield  {author} {\bibinfo {author} {\bibfnamefont {C.}~\bibnamefont
  {Wetterich}},\ }\bibfield  {title} {\bibinfo {title} {{Exact evolution
  equation for the effective potential}},\ }\href
  {https://doi.org/10.1016/0370-2693(93)90726-X} {\bibfield  {journal}
  {\bibinfo  {journal} {Phys. Lett. B}\ }\textbf {\bibinfo {volume} {301}},\
  \bibinfo {pages} {90} (\bibinfo {year} {1993})}\BibitemShut {NoStop}%
\bibitem [{\citenamefont {Kemler}\ and\ \citenamefont {Braun}(2013)}]{kem13}%
  \BibitemOpen
  \bibfield  {author} {\bibinfo {author} {\bibfnamefont {S.}~\bibnamefont
  {Kemler}}\ and\ \bibinfo {author} {\bibfnamefont {J.}~\bibnamefont {Braun}},\
  }\bibfield  {title} {\bibinfo {title} {{Towards a renormalization group
  approach to density functional theory--general formalism and case studies}},\
  }\href {https://doi.org/10.1088/0954-3899/40/8/085105} {\bibfield  {journal}
  {\bibinfo  {journal} {J. Phys. G}\ }\textbf {\bibinfo {volume} {40}},\
  \bibinfo {pages} {085105} (\bibinfo {year} {2013})}\BibitemShut {NoStop}%
\bibitem [{\citenamefont {Rentrop}\ \emph {et~al.}(2015)\citenamefont
  {Rentrop}, \citenamefont {Jakobs},\ and\ \citenamefont {Meden}}]{ren15}%
  \BibitemOpen
  \bibfield  {author} {\bibinfo {author} {\bibfnamefont {J.~F.}\ \bibnamefont
  {Rentrop}}, \bibinfo {author} {\bibfnamefont {S.~G.}\ \bibnamefont
  {Jakobs}},\ and\ \bibinfo {author} {\bibfnamefont {V.}~\bibnamefont
  {Meden}},\ }\bibfield  {title} {\bibinfo {title} {{Two-particle irreducible
  functional renormalization group schemes{\textemdash}a comparative study}},\
  }\href {https://doi.org/10.1088/1751-8113/48/14/145002} {\bibfield  {journal}
  {\bibinfo  {journal} {J. Phys. A}\ }\textbf {\bibinfo {volume} {48}},\
  \bibinfo {pages} {145002} (\bibinfo {year} {2015})}\BibitemShut {NoStop}%
\bibitem [{\citenamefont {Liang}\ \emph {et~al.}(2018)\citenamefont {Liang},
  \citenamefont {Niu},\ and\ \citenamefont {Hatsuda}}]{lia18}%
  \BibitemOpen
  \bibfield  {author} {\bibinfo {author} {\bibfnamefont {H.}~\bibnamefont
  {Liang}}, \bibinfo {author} {\bibfnamefont {Y.}~\bibnamefont {Niu}},\ and\
  \bibinfo {author} {\bibfnamefont {T.}~\bibnamefont {Hatsuda}},\ }\bibfield
  {title} {\bibinfo {title} {{Functional renormalization group and Kohn-Sham
  scheme in density functional theory}},\ }\href
  {https://doi.org/10.1016/j.physletb.2018.02.034} {\bibfield  {journal}
  {\bibinfo  {journal} {Phys. Lett. B}\ }\textbf {\bibinfo {volume} {779}},\
  \bibinfo {pages} {436} (\bibinfo {year} {2018})}\BibitemShut {NoStop}%
\bibitem [{\citenamefont {Kemler}\ \emph {et~al.}(2017)\citenamefont {Kemler},
  \citenamefont {Pospiech},\ and\ \citenamefont {Braun}}]{kem17a}%
  \BibitemOpen
  \bibfield  {author} {\bibinfo {author} {\bibfnamefont {S.}~\bibnamefont
  {Kemler}}, \bibinfo {author} {\bibfnamefont {M.}~\bibnamefont {Pospiech}},\
  and\ \bibinfo {author} {\bibfnamefont {J.}~\bibnamefont {Braun}},\ }\bibfield
   {title} {\bibinfo {title} {{Formation of selfbound states in a
  one-dimensional nuclear model--a renormalization group based density
  functional study}},\ }\href {https://doi.org/10.1088/0954-3899/44/1/015101}
  {\bibfield  {journal} {\bibinfo  {journal} {J. Phys. G}\ }\textbf {\bibinfo
  {volume} {44}},\ \bibinfo {pages} {015101} (\bibinfo {year}
  {2017})}\BibitemShut {NoStop}%
\bibitem [{\citenamefont {Yokota}\ \emph
  {et~al.}(2019{\natexlab{a}})\citenamefont {Yokota}, \citenamefont {Yoshida},\
  and\ \citenamefont {Kunihiro}}]{yok18}%
  \BibitemOpen
  \bibfield  {author} {\bibinfo {author} {\bibfnamefont {T.}~\bibnamefont
  {Yokota}}, \bibinfo {author} {\bibfnamefont {K.}~\bibnamefont {Yoshida}},\
  and\ \bibinfo {author} {\bibfnamefont {T.}~\bibnamefont {Kunihiro}},\
  }\bibfield  {title} {\bibinfo {title} {{Functional renormalization-group
  calculation of the equation of state of one-dimensional uniform matter
  inspired by the Hohenberg-Kohn theorem}},\ }\href
  {https://doi.org/10.1103/PhysRevC.99.024302} {\bibfield  {journal} {\bibinfo
  {journal} {Phys. Rev. C}\ }\textbf {\bibinfo {volume} {99}},\ \bibinfo
  {pages} {024302} (\bibinfo {year} {2019}{\natexlab{a}})}\BibitemShut
  {NoStop}%
\bibitem [{\citenamefont {Yokota}\ \emph
  {et~al.}(2019{\natexlab{b}})\citenamefont {Yokota}, \citenamefont {Yoshida},\
  and\ \citenamefont {Kunihiro}}]{yok18b}%
  \BibitemOpen
  \bibfield  {author} {\bibinfo {author} {\bibfnamefont {T.}~\bibnamefont
  {Yokota}}, \bibinfo {author} {\bibfnamefont {K.}~\bibnamefont {Yoshida}},\
  and\ \bibinfo {author} {\bibfnamefont {T.}~\bibnamefont {Kunihiro}},\
  }\bibfield  {title} {\bibinfo {title} {{\textit{Ab initio} description of
  excited states of 1D uniform matter with the Hohenberg--Kohn-theorem-inspired
  functional-renormalization-group method}},\ }\href
  {https://doi.org/10.1093/ptep/pty139} {\bibfield  {journal} {\bibinfo
  {journal} {Prog. Theor. Exp. Phys.}\ }\textbf {\bibinfo {volume} {2019}},\
  \bibinfo {pages} {011D01} (\bibinfo {year} {2019}{\natexlab{b}})}\BibitemShut
  {NoStop}%
\bibitem [{\citenamefont {Yokota}\ \emph {et~al.}(2021)\citenamefont {Yokota},
  \citenamefont {Haruyama},\ and\ \citenamefont {Sugino}}]{yok21b}%
  \BibitemOpen
  \bibfield  {author} {\bibinfo {author} {\bibfnamefont {T.}~\bibnamefont
  {Yokota}}, \bibinfo {author} {\bibfnamefont {J.}~\bibnamefont {Haruyama}},\
  and\ \bibinfo {author} {\bibfnamefont {O.}~\bibnamefont {Sugino}},\
  }\bibfield  {title} {\bibinfo {title} {{Functional-renormalization-group
  approach to classical liquids with short-range repulsion: A scheme without
  repulsive reference system}},\ }\href
  {https://doi.org/10.1103/PhysRevE.104.014124} {\bibfield  {journal} {\bibinfo
   {journal} {Phys. Rev. E}\ }\textbf {\bibinfo {volume} {104}},\ \bibinfo
  {pages} {014124} (\bibinfo {year} {2021})}\BibitemShut {NoStop}%
\bibitem [{\citenamefont {{Yokota}}\ and\ \citenamefont
  {{Naito}}(2019)}]{yok19}%
  \BibitemOpen
  \bibfield  {author} {\bibinfo {author} {\bibfnamefont {T.}~\bibnamefont
  {{Yokota}}}\ and\ \bibinfo {author} {\bibfnamefont {T.}~\bibnamefont
  {{Naito}}},\ }\bibfield  {title} {\bibinfo {title}
  {{Functional-renormalization-group aided density functional analysis for the
  correlation energy of the two-dimensional homogeneous electron gas}},\ }\href
  {https://doi.org/10.1103/PhysRevB.99.115106} {\bibfield  {journal} {\bibinfo
  {journal} {Phys. Rev. B}\ }\textbf {\bibinfo {volume} {99}},\ \bibinfo
  {pages} {115106} (\bibinfo {year} {2019})}\BibitemShut {NoStop}%
\bibitem [{\citenamefont {Yokota}\ and\ \citenamefont {Naito}(2021)}]{yok21}%
  \BibitemOpen
  \bibfield  {author} {\bibinfo {author} {\bibfnamefont {T.}~\bibnamefont
  {Yokota}}\ and\ \bibinfo {author} {\bibfnamefont {T.}~\bibnamefont {Naito}},\
  }\bibfield  {title} {\bibinfo {title} {{\textit{Ab initio} construction of
  the energy density functional for electron systems with the
  functional-renormalization-group-aided density functional theory}},\ }\href
  {https://doi.org/10.1103/PhysRevResearch.3.L012015} {\bibfield  {journal}
  {\bibinfo  {journal} {Phys. Rev. Research}\ }\textbf {\bibinfo {volume}
  {3}},\ \bibinfo {pages} {L012015} (\bibinfo {year} {2021})}\BibitemShut
  {NoStop}%
\bibitem [{\citenamefont {Martin}(2004)}]{Martin2004_CambridgeUniversityPress}%
  \BibitemOpen
  \bibfield  {author} {\bibinfo {author} {\bibfnamefont {R.~M.}\ \bibnamefont
  {Martin}},\ }\href@noop {} {\emph {\bibinfo {title} {{Electronic
  Structure}}}}\ (\bibinfo  {publisher} {Cambridge University Press},\ \bibinfo
  {year} {2004})\BibitemShut {NoStop}%
\bibitem [{\citenamefont {Jansen}(1991)}]{Jansen1991Phys.Rev.B43_12025}%
  \BibitemOpen
  \bibfield  {author} {\bibinfo {author} {\bibfnamefont {H.~J.~F.}\
  \bibnamefont {Jansen}},\ }\bibfield  {title} {\bibinfo {title} {{Many-body
  properties calculated from the Kohn-Sham equations in density-functional
  theory}},\ }\href {https://doi.org/10.1103/PhysRevB.43.12025} {\bibfield
  {journal} {\bibinfo  {journal} {Phys. Rev. B}\ }\textbf {\bibinfo {volume}
  {43}},\ \bibinfo {pages} {12025} (\bibinfo {year} {1991})}\BibitemShut
  {NoStop}%
\bibitem [{\citenamefont {von Barth}\ and\ \citenamefont
  {Hedin}(1972)}]{Barth1972J.Phys.C5_1629}%
  \BibitemOpen
  \bibfield  {author} {\bibinfo {author} {\bibfnamefont {U.}~\bibnamefont {von
  Barth}}\ and\ \bibinfo {author} {\bibfnamefont {L.}~\bibnamefont {Hedin}},\
  }\bibfield  {title} {\bibinfo {title} {{A local exchange-correlation
  potential for the spin polarized case. I}},\ }\href
  {https://doi.org/10.1088/0022-3719/5/13/012} {\bibfield  {journal} {\bibinfo
  {journal} {J. Phys. C}\ }\textbf {\bibinfo {volume} {5}},\ \bibinfo {pages}
  {1629} (\bibinfo {year} {1972})}\BibitemShut {NoStop}%
\bibitem [{\citenamefont {Akmal}\ \emph {et~al.}(1998)\citenamefont {Akmal},
  \citenamefont {Pandharipande},\ and\ \citenamefont
  {Ravenhall}}]{Akmal1998Phys.Rev.C58_1804}%
  \BibitemOpen
  \bibfield  {author} {\bibinfo {author} {\bibfnamefont {A.}~\bibnamefont
  {Akmal}}, \bibinfo {author} {\bibfnamefont {V.~R.}\ \bibnamefont
  {Pandharipande}},\ and\ \bibinfo {author} {\bibfnamefont {D.~G.}\
  \bibnamefont {Ravenhall}},\ }\bibfield  {title} {\bibinfo {title} {{Equation
  of state of nucleon matter and neutron star structure}},\ }\href
  {https://doi.org/10.1103/PhysRevC.58.1804} {\bibfield  {journal} {\bibinfo
  {journal} {Phys. Rev. C}\ }\textbf {\bibinfo {volume} {58}},\ \bibinfo
  {pages} {1804} (\bibinfo {year} {1998})}\BibitemShut {NoStop}%
\bibitem [{\citenamefont {Dickhoff}\ and\ \citenamefont
  {Barbieri}(2004)}]{Dickhoff2004Prog.Part.Nucl.Phys.52_377}%
  \BibitemOpen
  \bibfield  {author} {\bibinfo {author} {\bibfnamefont {W.}~\bibnamefont
  {Dickhoff}}\ and\ \bibinfo {author} {\bibfnamefont {C.}~\bibnamefont
  {Barbieri}},\ }\bibfield  {title} {\bibinfo {title} {{Self-consistent Green's
  function method for nuclei and nuclear matter}},\ }\href
  {https://doi.org/10.1016/j.ppnp.2004.02.038} {\bibfield  {journal} {\bibinfo
  {journal} {Prog. Part. Nucl. Phys.}\ }\textbf {\bibinfo {volume} {52}},\
  \bibinfo {pages} {377} (\bibinfo {year} {2004})}\BibitemShut {NoStop}%
\bibitem [{\citenamefont {Stone}\ and\ \citenamefont
  {Reinhard}(2007)}]{Stone2007Prog.Part.Nucl.Phys.58_587}%
  \BibitemOpen
  \bibfield  {author} {\bibinfo {author} {\bibfnamefont {J.~R.}\ \bibnamefont
  {Stone}}\ and\ \bibinfo {author} {\bibfnamefont {P.-G.}\ \bibnamefont
  {Reinhard}},\ }\bibfield  {title} {\bibinfo {title} {{The Skyrme interaction
  in finite nuclei and nuclear matter}},\ }\href
  {https://doi.org/10.1016/j.ppnp.2006.07.001} {\bibfield  {journal} {\bibinfo
  {journal} {Prog. Part. Nucl. Phys.}\ }\textbf {\bibinfo {volume} {58}},\
  \bibinfo {pages} {587} (\bibinfo {year} {2007})}\BibitemShut {NoStop}%
\bibitem [{\citenamefont {Gandolfi}\ \emph {et~al.}(2010)\citenamefont
  {Gandolfi}, \citenamefont {Illarionov}, \citenamefont {Fantoni},
  \citenamefont {Miller}, \citenamefont {Pederiva},\ and\ \citenamefont
  {Schmidt}}]{Gandolfi2010Mon.Not.R.Astron.Soc.404_L35}%
  \BibitemOpen
  \bibfield  {author} {\bibinfo {author} {\bibfnamefont {S.}~\bibnamefont
  {Gandolfi}}, \bibinfo {author} {\bibfnamefont {A.~Y.}\ \bibnamefont
  {Illarionov}}, \bibinfo {author} {\bibfnamefont {S.}~\bibnamefont {Fantoni}},
  \bibinfo {author} {\bibfnamefont {J.~C.}\ \bibnamefont {Miller}}, \bibinfo
  {author} {\bibfnamefont {F.}~\bibnamefont {Pederiva}},\ and\ \bibinfo
  {author} {\bibfnamefont {K.~E.}\ \bibnamefont {Schmidt}},\ }\bibfield
  {title} {\bibinfo {title} {{Microscopic calculation of the equation of state
  of nuclear matter and neutron star structure}},\ }\href
  {https://doi.org/10.1111/j.1745-3933.2010.00829.x} {\bibfield  {journal}
  {\bibinfo  {journal} {Mon. Not. R. Astron. Soc.}\ }\textbf {\bibinfo {volume}
  {404}},\ \bibinfo {pages} {L35} (\bibinfo {year} {2010})}\BibitemShut
  {NoStop}%
\bibitem [{\citenamefont {Lattimer}(2012)}]{lat12}%
  \BibitemOpen
  \bibfield  {author} {\bibinfo {author} {\bibfnamefont {J.~M.}\ \bibnamefont
  {Lattimer}},\ }\bibfield  {title} {\bibinfo {title} {{The Nuclear Equation of
  State and Neutron Star Masses}},\ }\href
  {https://doi.org/10.1146/annurev-nucl-102711-095018} {\bibfield  {journal}
  {\bibinfo  {journal} {Annu. Rev. Nucl. Part. Sci.}\ }\textbf {\bibinfo
  {volume} {62}},\ \bibinfo {pages} {485} (\bibinfo {year} {2012})}\BibitemShut
  {NoStop}%
\bibitem [{\citenamefont {Togashi}\ and\ \citenamefont
  {Takano}(2013)}]{Togashi2013Nucl.Phys.A902_53}%
  \BibitemOpen
  \bibfield  {author} {\bibinfo {author} {\bibfnamefont {H.}~\bibnamefont
  {Togashi}}\ and\ \bibinfo {author} {\bibfnamefont {M.}~\bibnamefont
  {Takano}},\ }\bibfield  {title} {\bibinfo {title} {{Variational study for the
  equation of state of asymmetric nuclear matter at finite temperatures}},\
  }\href {https://doi.org/10.1016/j.nuclphysa.2013.02.014} {\bibfield
  {journal} {\bibinfo  {journal} {Nucl. Phys. A}\ }\textbf {\bibinfo {volume}
  {902}},\ \bibinfo {pages} {53} (\bibinfo {year} {2013})}\BibitemShut
  {NoStop}%
\bibitem [{\citenamefont {Togashi}\ \emph {et~al.}(2016)\citenamefont
  {Togashi}, \citenamefont {Hiyama}, \citenamefont {Yamamoto},\ and\
  \citenamefont {Takano}}]{Togashi2016Phys.Rev.C93_035808}%
  \BibitemOpen
  \bibfield  {author} {\bibinfo {author} {\bibfnamefont {H.}~\bibnamefont
  {Togashi}}, \bibinfo {author} {\bibfnamefont {E.}~\bibnamefont {Hiyama}},
  \bibinfo {author} {\bibfnamefont {Y.}~\bibnamefont {Yamamoto}},\ and\
  \bibinfo {author} {\bibfnamefont {M.}~\bibnamefont {Takano}},\ }\bibfield
  {title} {\bibinfo {title} {{Equation of state for neutron stars with hyperons
  using a variational method}},\ }\href
  {https://doi.org/10.1103/PhysRevC.93.035808} {\bibfield  {journal} {\bibinfo
  {journal} {Phys. Rev. C}\ }\textbf {\bibinfo {volume} {93}},\ \bibinfo
  {pages} {035808} (\bibinfo {year} {2016})}\BibitemShut {NoStop}%
\bibitem [{\citenamefont {Oertel}\ \emph {et~al.}(2017)\citenamefont {Oertel},
  \citenamefont {Hempel}, \citenamefont {Kl\"ahn},\ and\ \citenamefont
  {Typel}}]{RevModPhys.89.015007}%
  \BibitemOpen
  \bibfield  {author} {\bibinfo {author} {\bibfnamefont {M.}~\bibnamefont
  {Oertel}}, \bibinfo {author} {\bibfnamefont {M.}~\bibnamefont {Hempel}},
  \bibinfo {author} {\bibfnamefont {T.}~\bibnamefont {Kl\"ahn}},\ and\ \bibinfo
  {author} {\bibfnamefont {S.}~\bibnamefont {Typel}},\ }\bibfield  {title}
  {\bibinfo {title} {{Equations of state for supernovae and compact stars}},\
  }\href {https://doi.org/10.1103/RevModPhys.89.015007} {\bibfield  {journal}
  {\bibinfo  {journal} {Rev. Mod. Phys.}\ }\textbf {\bibinfo {volume} {89}},\
  \bibinfo {pages} {015007} (\bibinfo {year} {2017})}\BibitemShut {NoStop}%
\bibitem [{\citenamefont {Tong}\ \emph {et~al.}(2018)\citenamefont {Tong},
  \citenamefont {Ren}, \citenamefont {Ring}, \citenamefont {Shen},
  \citenamefont {Wang},\ and\ \citenamefont
  {Meng}}]{Tong2018Phys.Rev.C98_054302}%
  \BibitemOpen
  \bibfield  {author} {\bibinfo {author} {\bibfnamefont {H.}~\bibnamefont
  {Tong}}, \bibinfo {author} {\bibfnamefont {X.-L.}\ \bibnamefont {Ren}},
  \bibinfo {author} {\bibfnamefont {P.}~\bibnamefont {Ring}}, \bibinfo {author}
  {\bibfnamefont {S.-H.}\ \bibnamefont {Shen}}, \bibinfo {author}
  {\bibfnamefont {S.-B.}\ \bibnamefont {Wang}},\ and\ \bibinfo {author}
  {\bibfnamefont {J.}~\bibnamefont {Meng}},\ }\bibfield  {title} {\bibinfo
  {title} {{Relativistic Brueckner-Hartree-Fock theory in nuclear matter
  without the average momentum approximation}},\ }\href
  {https://doi.org/10.1103/PhysRevC.98.054302} {\bibfield  {journal} {\bibinfo
  {journal} {Phys. Rev. C}\ }\textbf {\bibinfo {volume} {98}},\ \bibinfo
  {pages} {054302} (\bibinfo {year} {2018})}\BibitemShut {NoStop}%
\bibitem [{\citenamefont {Myo}\ \emph {et~al.}(2019)\citenamefont {Myo},
  \citenamefont {Takemoto}, \citenamefont {Lyu}, \citenamefont {Wan},
  \citenamefont {Xu}, \citenamefont {Toki}, \citenamefont {Horiuchi},
  \citenamefont {Yamada},\ and\ \citenamefont
  {Ikeda}}]{Myo2019Phys.Rev.C99_024312}%
  \BibitemOpen
  \bibfield  {author} {\bibinfo {author} {\bibfnamefont {T.}~\bibnamefont
  {Myo}}, \bibinfo {author} {\bibfnamefont {H.}~\bibnamefont {Takemoto}},
  \bibinfo {author} {\bibfnamefont {M.}~\bibnamefont {Lyu}}, \bibinfo {author}
  {\bibfnamefont {N.}~\bibnamefont {Wan}}, \bibinfo {author} {\bibfnamefont
  {C.}~\bibnamefont {Xu}}, \bibinfo {author} {\bibfnamefont {H.}~\bibnamefont
  {Toki}}, \bibinfo {author} {\bibfnamefont {H.}~\bibnamefont {Horiuchi}},
  \bibinfo {author} {\bibfnamefont {T.}~\bibnamefont {Yamada}},\ and\ \bibinfo
  {author} {\bibfnamefont {K.}~\bibnamefont {Ikeda}},\ }\bibfield  {title}
  {\bibinfo {title} {{Variational calculation of nuclear matter in a finite
  particle number approach using the unitary correlation operator and
  high-momentum pair methods}},\ }\href
  {https://doi.org/10.1103/PhysRevC.99.024312} {\bibfield  {journal} {\bibinfo
  {journal} {Phys. Rev. C}\ }\textbf {\bibinfo {volume} {99}},\ \bibinfo
  {pages} {024312} (\bibinfo {year} {2019})}\BibitemShut {NoStop}%
\bibitem [{\citenamefont {Wang}\ \emph {et~al.}(2021)\citenamefont {Wang},
  \citenamefont {Zhao}, \citenamefont {Ring},\ and\ \citenamefont
  {Meng}}]{Wang2021Phys.Rev.C103_054319}%
  \BibitemOpen
  \bibfield  {author} {\bibinfo {author} {\bibfnamefont {S.}~\bibnamefont
  {Wang}}, \bibinfo {author} {\bibfnamefont {Q.}~\bibnamefont {Zhao}}, \bibinfo
  {author} {\bibfnamefont {P.}~\bibnamefont {Ring}},\ and\ \bibinfo {author}
  {\bibfnamefont {J.}~\bibnamefont {Meng}},\ }\bibfield  {title} {\bibinfo
  {title} {{Nuclear matter in relativistic Brueckner-Hartree-Fock theory with
  Bonn potential in the full Dirac space}},\ }\href
  {https://doi.org/10.1103/PhysRevC.103.054319} {\bibfield  {journal} {\bibinfo
   {journal} {Phys. Rev. C}\ }\textbf {\bibinfo {volume} {103}},\ \bibinfo
  {pages} {054319} (\bibinfo {year} {2021})}\BibitemShut {NoStop}%
\bibitem [{\citenamefont {Yokota}\ \emph {et~al.}(2020)\citenamefont {Yokota},
  \citenamefont {Kasuya}, \citenamefont {Yoshida},\ and\ \citenamefont
  {Kunihiro}}]{yok20}%
  \BibitemOpen
  \bibfield  {author} {\bibinfo {author} {\bibfnamefont {T.}~\bibnamefont
  {Yokota}}, \bibinfo {author} {\bibfnamefont {H.}~\bibnamefont {Kasuya}},
  \bibinfo {author} {\bibfnamefont {K.}~\bibnamefont {Yoshida}},\ and\ \bibinfo
  {author} {\bibfnamefont {T.}~\bibnamefont {Kunihiro}},\ }\bibfield  {title}
  {\bibinfo {title} {{Microscopic derivation of density functional theory for
  superfluid systems based on effective action formalism}},\ }\href
  {https://doi.org/10.1093/ptep/ptaa173} {\bibfield  {journal} {\bibinfo
  {journal} {Prog. Theor. Exp. Phys.}\ }\textbf {\bibinfo {volume} {2021}},\
  \bibinfo {pages} {013A03} (\bibinfo {year} {2020})}\BibitemShut {NoStop}%
\bibitem [{\citenamefont {{Loos}}\ and\ \citenamefont {{Gill}}(2016)}]{loo16}%
  \BibitemOpen
  \bibfield  {author} {\bibinfo {author} {\bibfnamefont {P.-F.}\ \bibnamefont
  {{Loos}}}\ and\ \bibinfo {author} {\bibfnamefont {P.~M.~W.}\ \bibnamefont
  {{Gill}}},\ }\bibfield  {title} {\bibinfo {title} {{The uniform electron
  gas}},\ }\href {https://doi.org/https://doi.org/10.1002/wcms.1257} {\bibfield
   {journal} {\bibinfo  {journal} {WIREs Comput. Mol. Sci.}\ }\textbf {\bibinfo
  {volume} {6}},\ \bibinfo {pages} {410} (\bibinfo {year} {2016})}\BibitemShut
  {NoStop}%
\bibitem [{\citenamefont {Ceperley}\ and\ \citenamefont {Alder}(1980)}]{cep80}%
  \BibitemOpen
  \bibfield  {author} {\bibinfo {author} {\bibfnamefont {D.~M.}\ \bibnamefont
  {Ceperley}}\ and\ \bibinfo {author} {\bibfnamefont {B.~J.}\ \bibnamefont
  {Alder}},\ }\bibfield  {title} {\bibinfo {title} {{Ground State of the
  Electron Gas by a Stochastic Method}},\ }\href
  {https://doi.org/10.1103/PhysRevLett.45.566} {\bibfield  {journal} {\bibinfo
  {journal} {Phys. Rev. Lett.}\ }\textbf {\bibinfo {volume} {45}},\ \bibinfo
  {pages} {566} (\bibinfo {year} {1980})}\BibitemShut {NoStop}%
\bibitem [{\citenamefont {Vosko}\ \emph {et~al.}(1980)\citenamefont {Vosko},
  \citenamefont {Wilk},\ and\ \citenamefont {Nusair}}]{vos80}%
  \BibitemOpen
  \bibfield  {author} {\bibinfo {author} {\bibfnamefont {S.~H.}\ \bibnamefont
  {Vosko}}, \bibinfo {author} {\bibfnamefont {L.}~\bibnamefont {Wilk}},\ and\
  \bibinfo {author} {\bibfnamefont {M.}~\bibnamefont {Nusair}},\ }\bibfield
  {title} {\bibinfo {title} {{Accurate spin-dependent electron liquid
  correlation energies for local spin density calculations: a critical
  analysis}},\ }\href {https://doi.org/10.1139/p80-159} {\bibfield  {journal}
  {\bibinfo  {journal} {Can. J. Phys.}\ }\textbf {\bibinfo {volume} {58}},\
  \bibinfo {pages} {1200} (\bibinfo {year} {1980})}\BibitemShut {NoStop}%
\bibitem [{\citenamefont {Perdew}\ and\ \citenamefont {Zunger}(1981)}]{per81}%
  \BibitemOpen
  \bibfield  {author} {\bibinfo {author} {\bibfnamefont {J.~P.}\ \bibnamefont
  {Perdew}}\ and\ \bibinfo {author} {\bibfnamefont {A.}~\bibnamefont
  {Zunger}},\ }\bibfield  {title} {\bibinfo {title} {{Self-interaction
  correction to density-functional approximations for many-electron systems}},\
  }\href {https://doi.org/10.1103/PhysRevB.23.5048} {\bibfield  {journal}
  {\bibinfo  {journal} {Phys. Rev. B}\ }\textbf {\bibinfo {volume} {23}},\
  \bibinfo {pages} {5048} (\bibinfo {year} {1981})}\BibitemShut {NoStop}%
\bibitem [{\citenamefont {Ceperley}(1978)}]{cep78}%
  \BibitemOpen
  \bibfield  {author} {\bibinfo {author} {\bibfnamefont {D.}~\bibnamefont
  {Ceperley}},\ }\bibfield  {title} {\bibinfo {title} {{Ground state of the
  fermion one-component plasma: A Monte Carlo study in two and three
  dimensions}},\ }\href {https://doi.org/10.1103/PhysRevB.18.3126} {\bibfield
  {journal} {\bibinfo  {journal} {Phys. Rev. B}\ }\textbf {\bibinfo {volume}
  {18}},\ \bibinfo {pages} {3126} (\bibinfo {year} {1978})}\BibitemShut
  {NoStop}%
\bibitem [{\citenamefont {Ortiz}\ and\ \citenamefont {Ballone}(1994)}]{ort94}%
  \BibitemOpen
  \bibfield  {author} {\bibinfo {author} {\bibfnamefont {G.}~\bibnamefont
  {Ortiz}}\ and\ \bibinfo {author} {\bibfnamefont {P.}~\bibnamefont
  {Ballone}},\ }\bibfield  {title} {\bibinfo {title} {{Correlation energy,
  structure factor, radial distribution function, and momentum distribution of
  the spin-polarized uniform electron gas}},\ }\href
  {https://doi.org/10.1103/PhysRevB.50.1391} {\bibfield  {journal} {\bibinfo
  {journal} {Phys. Rev. B}\ }\textbf {\bibinfo {volume} {50}},\ \bibinfo
  {pages} {1391} (\bibinfo {year} {1994})}\BibitemShut {NoStop}%
\bibitem [{\citenamefont {Ortiz}\ and\ \citenamefont {Ballone}(1997)}]{ort97}%
  \BibitemOpen
  \bibfield  {author} {\bibinfo {author} {\bibfnamefont {G.}~\bibnamefont
  {Ortiz}}\ and\ \bibinfo {author} {\bibfnamefont {P.}~\bibnamefont
  {Ballone}},\ }\bibfield  {title} {\bibinfo {title} {{Erratum: Correlation
  energy, structure factor, radial distribution function, and momentum
  distribution of the spin-polarized uniform electron gas [Phys. Rev. B 50,
  1391 (1994)]}},\ }\href {https://doi.org/10.1103/PhysRevB.56.9970} {\bibfield
   {journal} {\bibinfo  {journal} {Phys. Rev. B}\ }\textbf {\bibinfo {volume}
  {56}},\ \bibinfo {pages} {9970} (\bibinfo {year} {1997})}\BibitemShut
  {NoStop}%
\bibitem [{\citenamefont {Kwon}\ \emph {et~al.}(1998)\citenamefont {Kwon},
  \citenamefont {Ceperley},\ and\ \citenamefont {Martin}}]{kwo98}%
  \BibitemOpen
  \bibfield  {author} {\bibinfo {author} {\bibfnamefont {Y.}~\bibnamefont
  {Kwon}}, \bibinfo {author} {\bibfnamefont {D.~M.}\ \bibnamefont {Ceperley}},\
  and\ \bibinfo {author} {\bibfnamefont {R.~M.}\ \bibnamefont {Martin}},\
  }\bibfield  {title} {\bibinfo {title} {{Effects of backflow correlation in
  the three-dimensional electron gas: Quantum Monte Carlo study}},\ }\href
  {https://doi.org/10.1103/PhysRevB.58.6800} {\bibfield  {journal} {\bibinfo
  {journal} {Phys. Rev. B}\ }\textbf {\bibinfo {volume} {58}},\ \bibinfo
  {pages} {6800} (\bibinfo {year} {1998})}\BibitemShut {NoStop}%
\bibitem [{\citenamefont {Ortiz}\ \emph {et~al.}(1999)\citenamefont {Ortiz},
  \citenamefont {Harris},\ and\ \citenamefont {Ballone}}]{ort99}%
  \BibitemOpen
  \bibfield  {author} {\bibinfo {author} {\bibfnamefont {G.}~\bibnamefont
  {Ortiz}}, \bibinfo {author} {\bibfnamefont {M.}~\bibnamefont {Harris}},\ and\
  \bibinfo {author} {\bibfnamefont {P.}~\bibnamefont {Ballone}},\ }\bibfield
  {title} {\bibinfo {title} {{Zero Temperature Phases of the Electron Gas}},\
  }\href {https://doi.org/10.1103/PhysRevLett.82.5317} {\bibfield  {journal}
  {\bibinfo  {journal} {Phys. Rev. Lett.}\ }\textbf {\bibinfo {volume} {82}},\
  \bibinfo {pages} {5317} (\bibinfo {year} {1999})}\BibitemShut {NoStop}%
\bibitem [{\citenamefont {Zong}\ \emph {et~al.}(2002)\citenamefont {Zong},
  \citenamefont {Lin},\ and\ \citenamefont {Ceperley}}]{zon02}%
  \BibitemOpen
  \bibfield  {author} {\bibinfo {author} {\bibfnamefont {F.~H.}\ \bibnamefont
  {Zong}}, \bibinfo {author} {\bibfnamefont {C.}~\bibnamefont {Lin}},\ and\
  \bibinfo {author} {\bibfnamefont {D.~M.}\ \bibnamefont {Ceperley}},\
  }\bibfield  {title} {\bibinfo {title} {{Spin polarization of the low-density
  three-dimensional electron gas}},\ }\href
  {https://doi.org/10.1103/PhysRevE.66.036703} {\bibfield  {journal} {\bibinfo
  {journal} {Phys. Rev. E}\ }\textbf {\bibinfo {volume} {66}},\ \bibinfo
  {pages} {036703} (\bibinfo {year} {2002})}\BibitemShut {NoStop}%
\bibitem [{\citenamefont {Drummond}\ \emph {et~al.}(2004)\citenamefont
  {Drummond}, \citenamefont {Towler},\ and\ \citenamefont {Needs}}]{dru04}%
  \BibitemOpen
  \bibfield  {author} {\bibinfo {author} {\bibfnamefont {N.~D.}\ \bibnamefont
  {Drummond}}, \bibinfo {author} {\bibfnamefont {M.~D.}\ \bibnamefont
  {Towler}},\ and\ \bibinfo {author} {\bibfnamefont {R.~J.}\ \bibnamefont
  {Needs}},\ }\bibfield  {title} {\bibinfo {title} {{Jastrow correlation factor
  for atoms, molecules, and solids}},\ }\href
  {https://doi.org/10.1103/PhysRevB.70.235119} {\bibfield  {journal} {\bibinfo
  {journal} {Phys. Rev. B}\ }\textbf {\bibinfo {volume} {70}},\ \bibinfo
  {pages} {235119} (\bibinfo {year} {2004})}\BibitemShut {NoStop}%
\bibitem [{\citenamefont {Spink}\ \emph {et~al.}(2013)\citenamefont {Spink},
  \citenamefont {Needs},\ and\ \citenamefont {Drummond}}]{spi13}%
  \BibitemOpen
  \bibfield  {author} {\bibinfo {author} {\bibfnamefont {G.~G.}\ \bibnamefont
  {Spink}}, \bibinfo {author} {\bibfnamefont {R.~J.}\ \bibnamefont {Needs}},\
  and\ \bibinfo {author} {\bibfnamefont {N.~D.}\ \bibnamefont {Drummond}},\
  }\bibfield  {title} {\bibinfo {title} {{Quantum Monte Carlo study of the
  three-dimensional spin-polarized homogeneous electron gas}},\ }\href
  {https://doi.org/10.1103/PhysRevB.88.085121} {\bibfield  {journal} {\bibinfo
  {journal} {Phys. Rev. B}\ }\textbf {\bibinfo {volume} {88}},\ \bibinfo
  {pages} {085121} (\bibinfo {year} {2013})}\BibitemShut {NoStop}%
\bibitem [{\citenamefont {Tanatar}\ and\ \citenamefont
  {Ceperley}(1989)}]{tan89}%
  \BibitemOpen
  \bibfield  {author} {\bibinfo {author} {\bibfnamefont {B.}~\bibnamefont
  {Tanatar}}\ and\ \bibinfo {author} {\bibfnamefont {D.~M.}\ \bibnamefont
  {Ceperley}},\ }\bibfield  {title} {\bibinfo {title} {{Ground state of the
  two-dimensional electron gas}},\ }\href
  {https://doi.org/10.1103/PhysRevB.39.5005} {\bibfield  {journal} {\bibinfo
  {journal} {Phys. Rev. B}\ }\textbf {\bibinfo {volume} {39}},\ \bibinfo
  {pages} {5005} (\bibinfo {year} {1989})}\BibitemShut {NoStop}%
\bibitem [{\citenamefont {Kwon}\ \emph {et~al.}(1993)\citenamefont {Kwon},
  \citenamefont {Ceperley},\ and\ \citenamefont {Martin}}]{kwo93}%
  \BibitemOpen
  \bibfield  {author} {\bibinfo {author} {\bibfnamefont {Y.}~\bibnamefont
  {Kwon}}, \bibinfo {author} {\bibfnamefont {D.~M.}\ \bibnamefont {Ceperley}},\
  and\ \bibinfo {author} {\bibfnamefont {R.~M.}\ \bibnamefont {Martin}},\
  }\bibfield  {title} {\bibinfo {title} {{Effects of three-body and backflow
  correlations in the two-dimensional electron gas}},\ }\href
  {https://doi.org/10.1103/PhysRevB.48.12037} {\bibfield  {journal} {\bibinfo
  {journal} {Phys. Rev. B}\ }\textbf {\bibinfo {volume} {48}},\ \bibinfo
  {pages} {12037} (\bibinfo {year} {1993})}\BibitemShut {NoStop}%
\bibitem [{\citenamefont {Rapisarda}\ and\ \citenamefont
  {Senatore}(1996)}]{rap96}%
  \BibitemOpen
  \bibfield  {author} {\bibinfo {author} {\bibfnamefont {F.}~\bibnamefont
  {Rapisarda}}\ and\ \bibinfo {author} {\bibfnamefont {G.}~\bibnamefont
  {Senatore}},\ }\bibfield  {title} {\bibinfo {title} {{Diffusion Monte Carlo
  study of electrons in two-dimensional layers}},\ }\href@noop {} {\bibfield
  {journal} {\bibinfo  {journal} {Aust. J. Phys.}\ }\textbf {\bibinfo {volume}
  {49}},\ \bibinfo {pages} {161} (\bibinfo {year} {1996})}\BibitemShut
  {NoStop}%
\bibitem [{\citenamefont {Attaccalite}\ \emph {et~al.}(2002)\citenamefont
  {Attaccalite}, \citenamefont {Moroni}, \citenamefont {Gori-Giorgi},\ and\
  \citenamefont {Bachelet}}]{att02}%
  \BibitemOpen
  \bibfield  {author} {\bibinfo {author} {\bibfnamefont {C.}~\bibnamefont
  {Attaccalite}}, \bibinfo {author} {\bibfnamefont {S.}~\bibnamefont {Moroni}},
  \bibinfo {author} {\bibfnamefont {P.}~\bibnamefont {Gori-Giorgi}},\ and\
  \bibinfo {author} {\bibfnamefont {G.~B.}\ \bibnamefont {Bachelet}},\
  }\bibfield  {title} {\bibinfo {title} {{Correlation Energy and Spin
  Polarization in the 2D Electron Gas}},\ }\href
  {https://doi.org/10.1103/PhysRevLett.88.256601} {\bibfield  {journal}
  {\bibinfo  {journal} {Phys. Rev. Lett.}\ }\textbf {\bibinfo {volume} {88}},\
  \bibinfo {pages} {256601} (\bibinfo {year} {2002})}\BibitemShut {NoStop}%
\bibitem [{\citenamefont {Attaccalite}\ \emph {et~al.}(2003)\citenamefont
  {Attaccalite}, \citenamefont {Moroni}, \citenamefont {Gori-Giorgi},\ and\
  \citenamefont {Bachelet}}]{att03}%
  \BibitemOpen
  \bibfield  {author} {\bibinfo {author} {\bibfnamefont {C.}~\bibnamefont
  {Attaccalite}}, \bibinfo {author} {\bibfnamefont {S.}~\bibnamefont {Moroni}},
  \bibinfo {author} {\bibfnamefont {P.}~\bibnamefont {Gori-Giorgi}},\ and\
  \bibinfo {author} {\bibfnamefont {G.~B.}\ \bibnamefont {Bachelet}},\
  }\bibfield  {title} {\bibinfo {title} {{Erratum: Correlation Energy and Spin
  Polarization in the 2D Electron Gas [Phys. Rev. Lett. 88, 256601 (2002)]}},\
  }\href {https://doi.org/10.1103/PhysRevLett.91.109902} {\bibfield  {journal}
  {\bibinfo  {journal} {Phys. Rev. Lett.}\ }\textbf {\bibinfo {volume} {91}},\
  \bibinfo {pages} {109902} (\bibinfo {year} {2003})}\BibitemShut {NoStop}%
\bibitem [{\citenamefont {Gori-Giorgi}\ \emph {et~al.}(2003)\citenamefont
  {Gori-Giorgi}, \citenamefont {Attaccalite}, \citenamefont {Moroni},\ and\
  \citenamefont {Bachelet}}]{gor03}%
  \BibitemOpen
  \bibfield  {author} {\bibinfo {author} {\bibfnamefont {P.}~\bibnamefont
  {Gori-Giorgi}}, \bibinfo {author} {\bibfnamefont {C.}~\bibnamefont
  {Attaccalite}}, \bibinfo {author} {\bibfnamefont {S.}~\bibnamefont
  {Moroni}},\ and\ \bibinfo {author} {\bibfnamefont {G.~B.}\ \bibnamefont
  {Bachelet}},\ }\bibfield  {title} {\bibinfo {title} {{Two-dimensional
  electron gas: Correlation energy versus density and spin polarization}},\
  }\href {https://doi.org/https://doi.org/10.1002/qua.10416} {\bibfield
  {journal} {\bibinfo  {journal} {Int. J. Quantum Chem.}\ }\textbf {\bibinfo
  {volume} {91}},\ \bibinfo {pages} {126} (\bibinfo {year} {2003})}\BibitemShut
  {NoStop}%
\bibitem [{\citenamefont {Drummond}\ and\ \citenamefont {Needs}(2009)}]{dru09}%
  \BibitemOpen
  \bibfield  {author} {\bibinfo {author} {\bibfnamefont {N.~D.}\ \bibnamefont
  {Drummond}}\ and\ \bibinfo {author} {\bibfnamefont {R.~J.}\ \bibnamefont
  {Needs}},\ }\bibfield  {title} {\bibinfo {title} {{Phase Diagram of the
  Low-Density Two-Dimensional Homogeneous Electron Gas}},\ }\href
  {https://doi.org/10.1103/PhysRevLett.102.126402} {\bibfield  {journal}
  {\bibinfo  {journal} {Phys. Rev. Lett.}\ }\textbf {\bibinfo {volume} {102}},\
  \bibinfo {pages} {126402} (\bibinfo {year} {2009})}\BibitemShut {NoStop}%
\bibitem [{\citenamefont {Gell-Mann}\ and\ \citenamefont
  {Brueckner}(1957)}]{gel57}%
  \BibitemOpen
  \bibfield  {author} {\bibinfo {author} {\bibfnamefont {M.}~\bibnamefont
  {Gell-Mann}}\ and\ \bibinfo {author} {\bibfnamefont {K.~A.}\ \bibnamefont
  {Brueckner}},\ }\bibfield  {title} {\bibinfo {title} {{Correlation Energy of
  an Electron Gas at High Density}},\ }\href
  {https://doi.org/10.1103/PhysRev.106.364} {\bibfield  {journal} {\bibinfo
  {journal} {Phys. Rev.}\ }\textbf {\bibinfo {volume} {106}},\ \bibinfo {pages}
  {364} (\bibinfo {year} {1957})}\BibitemShut {NoStop}%
\bibitem [{\citenamefont {Rajagopal}\ and\ \citenamefont
  {Kimball}(1977)}]{raj77}%
  \BibitemOpen
  \bibfield  {author} {\bibinfo {author} {\bibfnamefont {A.~K.}\ \bibnamefont
  {Rajagopal}}\ and\ \bibinfo {author} {\bibfnamefont {J.~C.}\ \bibnamefont
  {Kimball}},\ }\bibfield  {title} {\bibinfo {title} {{Correlations in a
  two-dimensional electron system}},\ }\href
  {https://doi.org/10.1103/PhysRevB.15.2819} {\bibfield  {journal} {\bibinfo
  {journal} {Phys. Rev. B}\ }\textbf {\bibinfo {volume} {15}},\ \bibinfo
  {pages} {2819} (\bibinfo {year} {1977})}\BibitemShut {NoStop}%
\bibitem [{\citenamefont {Lewin}\ \emph {et~al.}(2019)\citenamefont {Lewin},
  \citenamefont {Lieb},\ and\ \citenamefont {Seiringer}}]{lew19}%
  \BibitemOpen
  \bibfield  {author} {\bibinfo {author} {\bibfnamefont {M.}~\bibnamefont
  {Lewin}}, \bibinfo {author} {\bibfnamefont {E.~H.}\ \bibnamefont {Lieb}},\
  and\ \bibinfo {author} {\bibfnamefont {R.}~\bibnamefont {Seiringer}},\
  }\bibfield  {title} {\bibinfo {title} {{The local density approximation in
  density functional theory}},\ }\href@noop {} {\bibfield  {journal} {\bibinfo
  {journal} {Pure Appl. Anal.}\ }\textbf {\bibinfo {volume} {2}},\ \bibinfo
  {pages} {35} (\bibinfo {year} {2019})}\BibitemShut {NoStop}%
\bibitem [{\citenamefont {Dirac}(1930)}]{dir30}%
  \BibitemOpen
  \bibfield  {author} {\bibinfo {author} {\bibfnamefont {P.~A.~M.}\
  \bibnamefont {Dirac}},\ }\bibfield  {title} {\bibinfo {title} {{Note on
  Exchange Phenomena in the Thomas Atom}},\ }\href
  {https://doi.org/10.1017/S0305004100016108} {\bibfield  {journal} {\bibinfo
  {journal} {Math. Proc. Cambridge Philos. Soc.}\ }\textbf {\bibinfo {volume}
  {26}},\ \bibinfo {pages} {376–385} (\bibinfo {year} {1930})}\BibitemShut
  {NoStop}%
\bibitem [{\citenamefont {Friesecke}(1997)}]{fri97}%
  \BibitemOpen
  \bibfield  {author} {\bibinfo {author} {\bibfnamefont {G.}~\bibnamefont
  {Friesecke}},\ }\bibfield  {title} {\bibinfo {title} {{Pair Correlations and
  Exchange Phenomena in the Free Electron Gas}},\ }\href
  {https://doi.org/10.1007/s002200050056} {\bibfield  {journal} {\bibinfo
  {journal} {Commun. Math. Phys.}\ }\textbf {\bibinfo {volume} {184}},\
  \bibinfo {pages} {143} (\bibinfo {year} {1997})}\BibitemShut {NoStop}%
\bibitem [{\citenamefont {Dupuis}\ \emph {et~al.}(2021)\citenamefont {Dupuis},
  \citenamefont {Canet}, \citenamefont {Eichhorn}, \citenamefont {Metzner},
  \citenamefont {Pawlowski}, \citenamefont {Tissier},\ and\ \citenamefont
  {Wschebor}}]{dup21}%
  \BibitemOpen
  \bibfield  {author} {\bibinfo {author} {\bibfnamefont {N.}~\bibnamefont
  {Dupuis}}, \bibinfo {author} {\bibfnamefont {L.}~\bibnamefont {Canet}},
  \bibinfo {author} {\bibfnamefont {A.}~\bibnamefont {Eichhorn}}, \bibinfo
  {author} {\bibfnamefont {W.}~\bibnamefont {Metzner}}, \bibinfo {author}
  {\bibfnamefont {J.}~\bibnamefont {Pawlowski}}, \bibinfo {author}
  {\bibfnamefont {M.}~\bibnamefont {Tissier}},\ and\ \bibinfo {author}
  {\bibfnamefont {N.}~\bibnamefont {Wschebor}},\ }\bibfield  {title} {\bibinfo
  {title} {{The nonperturbative functional renormalization group and its
  applications}},\ }\href@noop {} {\bibfield  {journal} {\bibinfo  {journal}
  {Phys. Rep.}\ }\textbf {\bibinfo {volume} {910}},\ \bibinfo {pages} {1}
  (\bibinfo {year} {2021})}\BibitemShut {NoStop}%
\bibitem [{\citenamefont {Lue}(2015)}]{lue15}%
  \BibitemOpen
  \bibfield  {author} {\bibinfo {author} {\bibfnamefont {L.}~\bibnamefont
  {Lue}},\ }\bibfield  {title} {\bibinfo {title} {{Application of the
  functional renormalization group method to classical free energy models}},\
  }\href@noop {} {\bibfield  {journal} {\bibinfo  {journal} {AIChE Journal}\
  }\textbf {\bibinfo {volume} {61}},\ \bibinfo {pages} {2985} (\bibinfo {year}
  {2015})}\BibitemShut {NoStop}%
\bibitem [{\citenamefont {Ramakrishnan}\ and\ \citenamefont
  {Yussouff}(1979)}]{ram79}%
  \BibitemOpen
  \bibfield  {author} {\bibinfo {author} {\bibfnamefont {T.~V.}\ \bibnamefont
  {Ramakrishnan}}\ and\ \bibinfo {author} {\bibfnamefont {M.}~\bibnamefont
  {Yussouff}},\ }\bibfield  {title} {\bibinfo {title} {{First-principles
  order-parameter theory of freezing}},\ }\href
  {https://doi.org/10.1103/PhysRevB.19.2775} {\bibfield  {journal} {\bibinfo
  {journal} {Phys. Rev. B}\ }\textbf {\bibinfo {volume} {19}},\ \bibinfo
  {pages} {2775} (\bibinfo {year} {1979})}\BibitemShut {NoStop}%
\end{thebibliography}

\end{document}